\documentclass[epj]{svjour}

\usepackage{graphicx,cite}

\newenvironment{eq}[1]%
{\begin{bequation}{#1}}{\end{bequation}}

\newenvironment{eqarray}[1]%
{\begin{beqnarray}{#1}}{\end{beqnarray}}

\def\eqref#1{(\ref{#1})}

\newcommand{\inverse}[1]{ {1  \over {#1}} }

\newcommand{\twid}{\sim}

\newcommand{\ddt}{ {d \over dt} }

\newcommand{\limzero}[1]{ \lim_{#1 \rightarrow 0} }
\newcommand{\liminfty}[1]{ \lim_{#1 \rightarrow \infty} }

\newcommand{\paren}[1]{\left( #1 \right)}
\newcommand{\square}[1]{\left[ #1 \right]}
\newcommand{\curly}[1]{ \left\{ #1 \right\} }

\newcommand{\casesbracketsshortii}[4]
{\left\{
\begin{array}{ll}
#1\ & (#2) \\  #3\ & (#4) 
\end{array}%
\right.
}

\newcommand{\casesbracketsshortiii}[6]{
\left\{
\begin{array}{lll} 
#1\  & (#2) \\  #3\ & (#4) \\  #5\ &(#6)   
\end{array}%
\right.
}

\newcommand{\gap}{\hspace{.4in}}

\newcommand{\blank}{\ \\}

\newcommand{\gt}{\rightarrow}

\newcommand{\period}{\ \ .}
\newcommand{\comma}{\ ,\ }

\newcommand{\lsim}{\,\stackrel{<}{\scriptstyle \sim}\,}

\newcommand{\ignore}[1]{}

{\end{list}}

{\end{list}}

{\ \\ XXXXX \hfill XXXXX #1 XXXXX\begin{equation}\label{#1}}%
{\end{equation}}

{\ \\ XXXXX \hfill XXXXX #1 XXXXX\begin{eqnarray}\label{#1}}%
{\end{eqnarray}}

\newenvironment{bequation}[1]%
{\begin{equation}\label{#1}}%
{\end{equation}}

\newenvironment{beqnarray}[1]%
{\begin{eqnarray}\label{#1}}%
{\end{eqnarray}}

\newcommand{\drop}{\nonumber \\}

\newcommand{\ie}{i.\,e.~}

{\begin{center} \section*{#1} \end{center}}

\newcommand{\degreesc}{$^0$C }

{\end{figure}} 

{\end{figure}}

{\end{figure}}

\newcommand{\phiinf}{\phi_{\infty}}
\newcommand{\vinf}{v_{\infty}}
\newcommand{\Dinf}{D_{\infty}}
\newcommand{\minf}{m_{\infty}}
\newcommand{\Nbar}{\bar{N}}
\newcommand{\Nbart}{\bar{N}_t}
\newcommand{\Nbarinf}{\bar{N}_{\infty}}
\newcommand{\Nbartsquared}{\overline{{N}_{t}^2}}
\newcommand{\phit}{\phi_t}
\newcommand{\vt}{v_t}
\newcommand{\Dt}{D_t}
\newcommand{\mt}{m_t}
\newcommand{\mtot}{m_{\rm tot}}
\newcommand{\jt}{j_t}

\newcommand{\kplus}{k^+}
\newcommand{\vminus}{v^-}

\newcommand{\tauqs}{\tau_{\rm qs}}

\newcommand{\tstar}{t^{*}}

\newcommand{\epsilonc}{\epsilon_{\rm c}}
\newcommand{\Gtran}{G^{\rm tran}}
\newcommand{\Gdiff}{G^{\rm diff}}
\newcommand{\Glinear}{G^{\rm linear}}
\newcommand{\erfc}{{\rm erfc}}
\newcommand{\erf}{{\rm erf}}
\newcommand{\St}{S_t}
\newcommand{\Sinf}{S_\infty}
\newcommand{\tcross}{t_{\rm cross}}
\newcommand{\Deltat}{\Delta_t}
\newcommand{\Deltainf}{\Delta_\infty}

\newcommand{\taufill}{\tau_{\rm fill}}
\newcommand{\taufast}{\tau_{\rm fast}}
\newcommand{\tauslow}{\tau_{\rm slow}}

\begin{document}

%\bibliographystyle{benaip}

%****************************** title page *****************************************
\title{Dynamics of Living Polymers}

\author{
Ben O'Shaughnessy \inst{1} \thanks{\email{bo8@columbia.edu}}
\and 
Dimitrios Vavylonis\inst{1,2} \thanks{\email{dv35@columbia.edu}}
%\thanks{\emph{Current address:} Department of
%Chemical Engineering, Columbia University, New York, NY 10027}
% etc
%\author{First author\inst{1} \and Second author\inst{2}% etc
}                     % Do not remove
\institute{
Department of Chemical Engineering, Columbia University, 
500 West 120th Street, New York, NY 10027, USA
 \and
Department of Physics, Columbia University, 
538 West 120th Street, New York, NY 10027, USA
}

\date{}
% Submitted to: {\em  European Physical Journal E}\\

%****************************** abstract *****************************************

\abstract{We study theoretically the dynamics of living polymers which
can add and subtract monomer units at their live chain ends.  The
classic example is ionic living polymerization. In equilibrium, a
delicate balance is maintained in which each initiated chain has a
very small negative average growth rate (``velocity'') just sufficient
to negate the effect of growth rate fluctuations. This leads to an
exponential molecular weight distribution (MWD) with mean $\Nbar$.
After a small perturbation of relative amplitude $\epsilon$, e.g. a
small temperature jump, this balance is destroyed: the velocity
acquires a boost greatly exceeding its tiny equilibrium value.  For
$\epsilon > \epsilonc \approx 1/\Nbar^{1/2}$ the response has 3
stages: (1) Coherent chain growth or shrinkage, leaving a highly
non-linear hole or peak in the MWD at small chain lengths.  During
this episode, lasting time $\taufast \twid \Nbar$, the MWD's first
moment and monomer concentration $m$ relax very close to
equilibrium. (2) Hole-filling (or peak decay) after $\taufill \twid
\epsilon^2 \Nbar^2$.  The absence or surfeit of small chains is
erased. (3) Global MWD shape relaxation after $\tauslow \sim \Nbar^2$.
By this time second and higher MWD moments have relaxed. During
episodes (2) and (3) the fast variables ($\Nbar,m$) are enslaved to
the slowly varying number of free initiators (chains of zero length).
Thus fast variables are quasi-statically fine-tuned to equilibrium.
The outstanding feature of these dynamics is their ultrasensitivity:
despite the perturbation's linearity, the response is non-linear until
the late episode (3).  For very small perturbations, $\epsilon <
\epsilonc$, response remains non-linear but with a less dramatic peak
or hole during episode (1).  Our predictions are in agreement with
viscosity measurements on the most widely studied system,
$\alpha$-methylstyrene.}

\PACS{
    {82.35.-x} {Polymers: properties; reactions; polymerization.}      
    {05.40.-a} {Fluctuation phenomena, random processes, noise, and Brownian Motion.}
    {87.15.Rn} {Biomolecules: structure and physical properties; Reactions and kinetics; polymerization.}
} % end of PACS codes

\maketitle
%

%****************************** PAPER  *****************************************

\section{Introduction}

Living polymers are intriguing examples of soft matter and have major
roles in technology and biology. The classic system, of major
importance for the synthesis of high-grade polymer materials and the
subject of a long history of fundamental experimental and theoretical
study, is living anionic polymerization
\cite{szwarcvanbeylen:book,greer:review,greer:annual_review,greer:living_review_jphyschem,%
zheng:greer:living_1,zheng:greer:living_2,andrews:greer:living_3,%
sarkar:greer:living_4,zhuang:greer:living_5,pfeuty:greer:living_preliminary,%
garcia:greer:living_nonumber,zhuang:greer_kinetics,das:greer:living_7_mwd,%
kennedywheeler:living_critical,wheeler:polymerization_critical,%
wheelerpfeuty:polymerization,wheelerpfeuty:polymerization_fluc,%
schafer:rg_mwd,kaufman:plztn_equiv_neighbor_lattice,%
wheelerpetschek:living_mwd_bound,dudowicz:living_i,dudowicz:living_ii,%
dudowicz:living_iii,schoot:living_scaling,%
frischknechtmilner:living_aggregation}.  This class of polymerization
is widely employed to manufacture polymers with nearly monodisperse
MWDs and controlled architectures such as block and star copolymers
\cite{webster:living_review}.  A recent variant on this theme with
immense potential technological impact is living free radical
polymerization \cite{hawker:review:living_frp,georges:living_frp}.
Other examples include worm-like surfactant micelles
\cite{marques:end-evaporation,milchev:monomer_mediated_relaxation,%
marquescates:wlike_dynamics_exact,catescandau:living_review,%
wittmer:milchev:cates:equilibrium_polymers_montecarlo,%
wittmer:wormlike_mwd_numerical,spenley:rheology_wormlike,ben:living_letter}
and biological polymers such as actin and tubulin filaments
\cite{oosawaasakura:book,hill:aggr_bio_book,korn:actin_review_science,%
pollardcooper:actin_review,desaimitchison:mtub_review,%
flyvbjerg:leibler_microtubules_prl} whose
special properties are exploited by living cells for motility and
structural integrity.

As polymeric or polymer-like materials, the unique feature of these
systems is that the chains are dynamic objects, with constantly
fluctuating lengths.  When subjected to an external perturbation, they
are able to respond dynamically via polymerization and
depolymerization reactions allowing a new thermodynamic equilibrium to
be attained. These systems are ``living'' in the sense that a change
in their environment leads to a new equilibrium molecular weight
distribution (MWD).  All of this should be compared to conventional
inert polymer materials whose MWDs are frozen.

The subject of this paper is the dynamical response of living
polymers. We study the class of living polymers which is exemplified
by living ionic polymerization, having the following characteristics:
(i) monomers add and subtract from chain ends and (ii) the total
number of chains is fixed.  In the biological world, actin and
microtubule filaments satisfy condition (i), and the dynamics of both
filaments and filament caps are frequently described by situation (ii)
\cite{ben:actin_cap_letter}.  Wormlike micelles on the other hand in
some cases grow from their ends only
\cite{marques:end-evaporation,milchev:monomer_mediated_relaxation}
(condition (i)) but do not satisfy condition (ii): the number of
micelles is not conserved.  Interestingly, this makes their dynamical
response very different (see discussion section).

\begin{figure}[tb]             
\centering
\includegraphics[width=8cm]{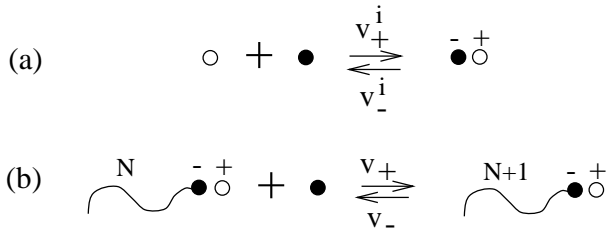}
\caption{%
(a) Schematic of initiation reaction in ionic polymerization. 
Empty (filled) circle represents initiator (monomer). 
(b) Polymerization and depolymerization reactions at the
ionic chain end of a polymer chain of $N$ units.}
\label{ionic_scheme}
\end{figure}

Our presentation will focus on living ionic polymerization, though
most of our results are completely general.  In these systems,
polymers grow reversibly from their ionic ends (see
fig. \ref{ionic_scheme}).  Due to their ``living'' nature, if fresh
monomer is added polymers will grow until the extra monomer has been
consumed; further addition of a different monomer species, say, will
resume polymer growth.  The main achievement of this technique is a
distribution of chain lengths with very small Poisson fluctuations
around the mean.  This monodisperse MWD is not however a true
equilibrium distribution.  For sufficiently long times slow
depolymerization reactions become important and lead to a true
equilibrium MWD which is in fact extremely broad.  For many
applications, for example styrene, the timescale to reach equilibrium
is extremely large and depolymeri\-zation-induced broadening effects are
unimportant \cite{miyakestockmayer:living}.  However, in some cases
equilibration times are accessible, an example being
$\alpha$-methylstyrene whose equilibrium properties have been studied
in many experiments
\cite{greer:review,greer:annual_review,greer:living_review_jphyschem,%
zheng:greer:living_1,zheng:greer:living_2,andrews:greer:living_3,%
sarkar:greer:living_4,zhuang:greer:living_5,pfeuty:greer:living_preliminary,%
garcia:greer:living_nonumber,zhuang:greer_kinetics,das:greer:living_7_mwd}.

%%%%%%%%%%%%%%%%%%%%%%%%%%%%%%%%%%%%%%%%%%%%%%%%%%%%%%%%%%%%%%%%
%%%%%%%%%%%%%%%%%%%%%%%%%%%%%%%%%%%%%%%%%%%%%%%%%%%%%%%%%%%%%%%%

\begin{figure}[bt]             
\centering
\includegraphics[width=8.5cm]{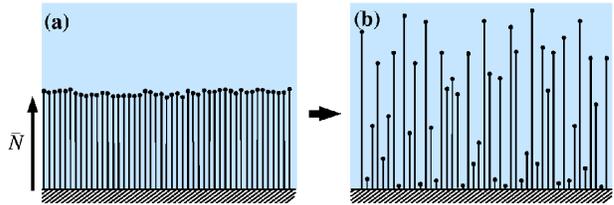}
\caption
{   Fluctuations of
living polymer chain length in equilibrium are very large.  For
clarity, living chains are depicted as if extending from an imaginary substrate.
Living ends are shown as filled circles.
Even if one starts with a nearly monodisperse MWD as in (a)
and with monomer concentration (grey background) such that
polymerization and depolymerization rates balance, random
polymerization and depolymerization reactions eventually broaden the
distribution up to distances of the same order as the mean $\Nbar$, as
shown in (b).  Note the resulting MWD includes chains of ``zero'' length
(free initiators).  }
\label{mono_to_broad}
\end{figure}

The reason that chain length fluctuations in equilibrium are so large
is that even when a state is reached where the monomer concentration
reaches a steady state value such that polymer growth rates exactly
balance depolymerization, through polymerization and depolymerization
reactions monomers continue to be reshuffled among living chains.
Thus even if one starts with an essentially monodisperse MWD as shown
in fig. \ref{mono_to_broad}(a), random addition and subtraction of
monomers from chain ends will result in a ``diffusive'' random walk
motion of chain ends in $N$-space (where $N$ is the number of monomer
units added to initiator) with a certain characteristic
``diffusivity'' $D$.  Since there is no restoring force to this
motion, chain ends will eventually diffuse to distances of the same
order as the mean chain length, resulting in a broad distribution as
shown in fig. \ref{mono_to_broad}(b).  In a simple mean-field picture
where monomer-monomer excluded volume interactions are neglected this
leads to the Flory-Schultz equilibrium distribution
\cite{flory:book,tobolsky:living_mwd,greer:annual_review}
%_______________________________________________________________________
                                                \begin{eq}{flory}
\phiinf(N) = \inverse{\Nbarinf} e^{-N/\Nbarinf} \comma
                                                                \end{eq}
%-----------------------------------------------------------------------
where $\Nbarinf$ is the equilibrium mean length.

%%%%%%%%%%%%%%%%%%%%%%%%%%%%%%%%%%%%%%%%%%%%%%%%%%%%%%%%%%%%%%%%
%%%%%%%%%%%%%%%%%%%%%%%%%%%%%%%%%%%%%%%%%%%%%%%%%%%%%%%%%%%%%%%%

The thermodynamic properties of equilibrium living polymers have been
studied in a large number of experimental and theoretical works.
According to theory, going beyond mean field by accounting for
excluded-volume interactions leads to power law modifications
\cite{schafer:rg_mwd,kaufman:plztn_equiv_neighbor_lattice,%
wheelerpetschek:living_mwd_bound,wittmer:wormlike_mwd_numerical} to
eq. \eqref{flory}.  Moreover, an analogy
\cite{kennedywheeler:living_critical,wheeler:polymerization_critical,%
wheelerpfeuty:polymerization,wheelerpfeuty:polymerization_fluc}
between a living polymer system and the much studied Ising spin system
in the formal limit of spin dimensionality $n$ becoming zero (similar
to the well-known analogy in the thermodynamic theory of semidilute
polymer solutions \cite{gennes:book,cloizeauxjannink:book} which
can in fact be formulated equivalently to a living polymer system
\cite{cloizeaux:pol_rg_many}) showed that the polymerization phase
transition near the polymerization temperature is second-order with a
number of characteristic power-law exponents.  The results of a large
number of experiments by the group of Greer, measuring both the MWD
itself \cite{das:greer:living_7_mwd} and other thermodynamic
properties \cite{greer:review,greer:living_review_jphyschem} are
apparently consistent with the above theories, favoring somewhat the
non mean-field approach.

\begin{figure}[tb]             
\centering
\includegraphics[width=8cm]{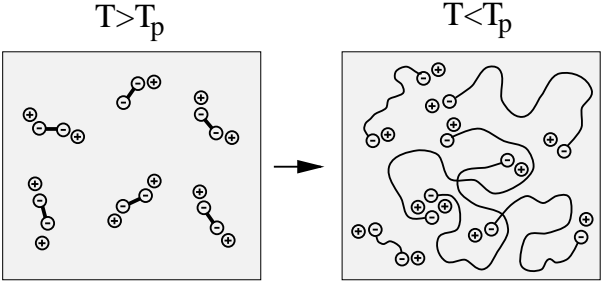}
\caption[$T$-quench Experiment of Greer et al.]
{Schematic of one of the classes of experiment performed by Greer's
group \cite{zhuang:greer_kinetics,das:greer:living_7_mwd}.
The mix of monomer, initiator and solvent is stored for sufficiently
long above the polymerization transition temperature $T_p$ such that
all initiator is activated. At this point all living chains are of
``zero'' length; we refer to these as ``free initiators'' 
(bifunctional dimers in this case).  After a quench below
$T_p$, polymerization onsets and the system's response to abrupt
temperature changes is then studied.  In this paper we study 
more elementary perturbations: very small $T$-jumps of an already equilibrated
system.}
\label{greer_quench}
\end{figure}

The most essential feature of living polymers however, namely their
ability to respond {\em dynamically} to external perturbations,
remains largely unexplored.  How does an equilibrium living polymer
system respond to a small perturbation?  A conceptually simple
perturbation is the addition of a small amount of extra monomer (an
``$m$-jump'').  A similar perturbation, easier to perform in practice,
is a ``$T$-jump,'' \ie a small sudden change in temperature.  Under
such a perturbation, how rapidly and in what manner will the monomer
concentration, MWD and mean chain length reach a new equilibrium?  In
this paper we address these questions theoretically.  We will compare
our predictions in section 8 with small $T$-jump viscosity relaxation
experiments by Greer et al. \cite{garcia:greer:living_nonumber}.
This same group has also monitored monomer and MWD dynamics following
much stronger perturbations, finite temperature
quenches\cite{zhuang:greer_kinetics,das:greer:living_7_mwd} (see
fig. \ref{greer_quench}).

%%%%%%%%%%%%%%%%%%%%%%%%%%%%%%%%%%%%%%%%%%%%%%%%%%%%%%%%%%%%%%%%
%%%%%%%%%%%%%%%%%%%%%%%%%%%%%%%%%%%%%%%%%%%%%%%%%%%%%%%%%%%%%%%%

Our main broad conclusion is that, counter-intuitively, the response to
a very small perturbation is extremely strong: even though the initial
and final MWDs (the latter attained after relaxation is complete) are
very close to one another, the MWD at intermediate times is very
strongly perturbed from equilibrium, \ie its shape is very different
from either the initial or final equilibrium shapes. In this sense,
there are no small perturbations: a small thermodynamic perturbation
well-described by a linear susceptibility (i.e. one inducing a small
change in the equilibrium state) has nonetheless a large dynamical
effect; intermediate states deviate strongly from equilibrium in the
sense that observables are perturbed in a highly non-linear manner. We
refer to this as dynamical ultrasensitivity.

%%%%%%%%%%%%%%%%%%%%%%%%%%%%%%%%%%%%%%%%%%%%%%%%%%%%%%%%%%%%%%%%
\begin{figure*}[htb]             
 \centering
\includegraphics[width=15cm]{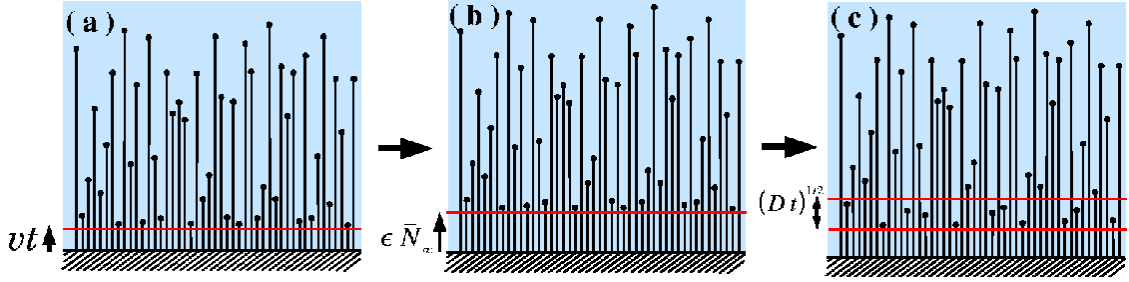}
\caption
{There are three stages in the relaxation of living polymers to
equilibrium. During the first, living polymers grow
coherently as in (a).  The translation stops when they have consumed enough
monomer such that monomer concentration drops to a value for
which polymerization and depolymerization balance (shown in (b)).
During the second stage, (c), through ``diffusive'' random
polymerization and depolymerization reactions, the MWD slowly fills
the depletion of chains of short lengths.  The third and final stage
entails global shape relaxation of the MWD on a diffusion timescale
corresponding to $\Nbarinf$.
}
\label{brush_stages}
\end{figure*}
%%%%%%%%%%%%%%%%%%%%%%%%%%%%%%%%%%%%%%%%%%%%%%%%%%%%%%%%%%%%%%%%

A hint of this sensitivity is already apparent in equilibrium, where a
very delicate balance is established between growth and shrinkage such
that the net polymerization velocity $v$ is essentially zero, of order
$1/\Nbarinf$ (due to the small fraction of special non-depolymerizable
chains of length $N=0$).  Now the effect of a small T-jump (say) is to
produce a small but finite velocity which overwhelms this tiny
velocity value and destroys the delicate balance sustaining
equilibrium.  Thus if for example $v$ becomes positive, the MWD will
{\em uniformly translate} toward larger molecular weights, creating a
depletion in the MWD at small chain lengths, as shown schematically in
fig \ref{brush_stages}.  This translation process stops only after
sufficient monomer has been polymerized such that the monomer
concentration drops close to its destination equilibrium value,
corresponding to a recovery of the zero net growth rate equilibrium
condition.  We find that only on a much longer timescale does
``diffusion'' of chain ends fill up the hole and lead the system to
the destination equilibrium MWD, as shown in fig. \ref{brush_stages}.

Why does this hole develop in the MWD?  The reason is that chain
length scales homogenized by diffusion, $(D t)^{1/2}$, are smaller
than scales affected by translational motion, $v t$, for times longer
than the timescale $\tstar \approx D/v^2$.  Thus if $\tstar$ is
shorter than the timescale needed for translation to stop, a depletion
hole develops in the MWD.  But even when translation is completed
earlier than $\tstar$, we find that the perturbation is still
nonlinear; even though no hole is formed there is still a large
amplitude reduction in the MWD, much larger than the relative
magnitude of the initial perturbation, $\epsilon$.

\begin{figure}[tb]             
\centering
\includegraphics[width=8.3cm]{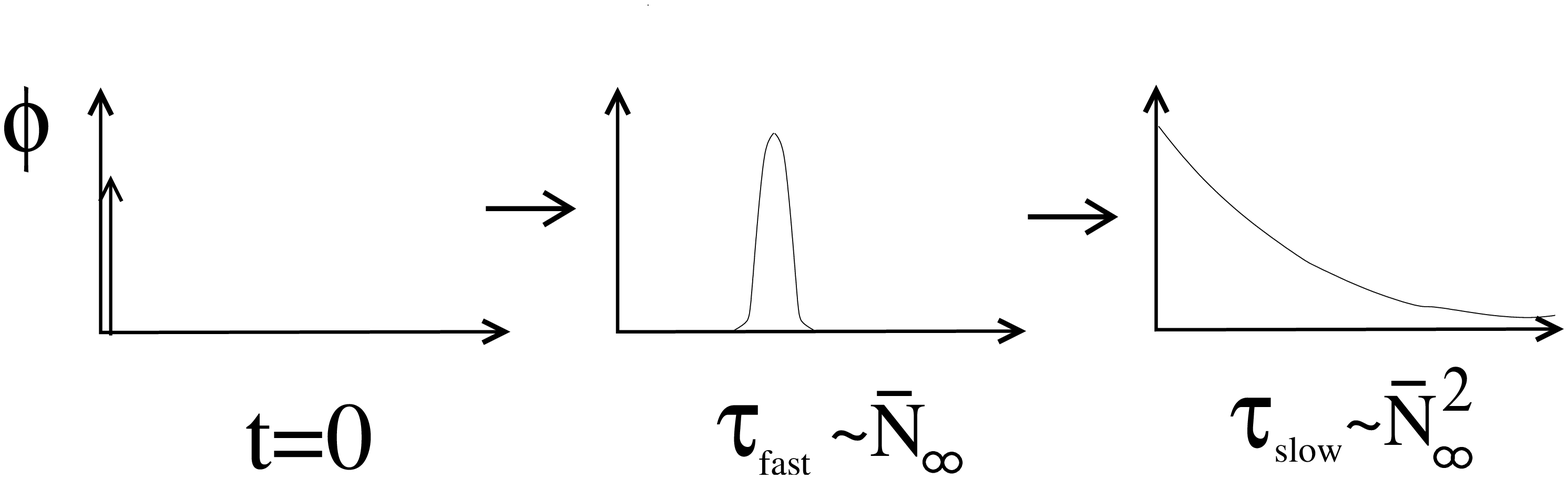}
\caption{Schematic of the MWD dynamics as analyzed by Miyake and Stockmayer 
\cite{miyakestockmayer:living} and their theoretical predictions regarding
the timescales involved.  (a) Initially ($t=0$) all initiators
are unpolymerized. (b) For $t \ll \taufast$ chains
grow coherently with a narrow MWD until its peak reaches
the equilibrium mean length $\Nbarinf$ at times of order $\taufast$. 
(c) During a third stage lasting up to $\tauslow$, which 
was beyond their theoretical analysis, the MWD spreads to the
equilibrium MWD.}
\label{miyake_stages}
\end{figure}

This work is related to earlier theory of Miyake and Stockmayer
\cite{miyakestockmayer:living} who, following a prior treatment by
Brown and Szwarc \cite{brownszwarc:living_ancient}, studied
analytically and numerically the dynamics in the special case where
initially all monomer is unpolymerized (see fig. \ref{miyake_stages}).
Their analytical results, applicable in the case of very small
depolymerization rates, showed a 2-stage relaxation.  During the
first, lasting a time $\twid \Nbarinf$, living chains grow coherently
from zero length up to the equilibrium value.  In the second stage the
shape of the sharply peaked MWD relaxes slowly toward the equilibrium
exponential distribution.  Although their perturbation analysis broke
down during this stage, Miyake and Stockmayer estimated that its
duration scales as $\Nbarinf^2$.  Later theoretical studies addressed
the second stage dynamics
\cite{nandajain:living_broadening,auluck:living_mwd_intermediate,taganov:living_mwd}
of this special case.

A short announcement of the present work has already appeared
\cite{ben:living_ionic_letter}.  The structure of the paper is as
follows.  In section 2 we establish the dynamical equations obeyed by
living polymers which show the nonlinear coupling between the monomer
population and the living chain MWD.  For simplicity, we neglect
excluded volume interactions and chain length dependent polymerization
rates.  Starting from the equations of section 2 we then analyze the
response to a $T$-jump. This response depends on the sign of the
velocity induced by the perturbation.  The case of positive initial
velocity is analyzed in sections 3 and 4, while the negative in
section 5.  The special case of very small $T$-jumps is analyzed in
section 6.  In section 7 we show that the results of the previous
sections directly generalize to small perturbations of arbitrary form.
Finally we conclude with a discussion of the results, and the
experimental outlook.

%************************************************************************************
%************************************************************************************
%************************************************************************************
%************************************************************************************
  
\section{Living Polymer Dynamics: Monomer-Polymer Coupling}

An important aspect of living polymers dynamics is the coupling
between the two distinct species present in the solution, free
monomers and living chain MWD: on the one hand living chain growth
rates depend on the concentration of free monomer and on the other
hand, due to mass conservation, monomer concentration if a function of
the living chain MWD.  Stated differently, living chains grow
according to a velocity field which is self-consistently updated
depending on the response of the MWD.  In this section we develop
the equations obeyed by monomer and MWD.  The dynamical response
to perturbations is discussed in the following sections.

A crucial feature of the living polymer systems we study is that the
number of living chains remains fixed.  This is what arises in the
experiments of Greer et
al. \cite{greer:review,greer:living_review_jphyschem,greer:annual_review},
where this number is determined by the amount of initiator initially
added in the solution.  Denoting $i_0$ the concentration of living
chains (\ie chains which have been initiated), mass conservation
implies that monomer concentration, $\mt$, and MWD, $\phit(N)$, ($t$
denotes time) obey
%_______________________________________________________________________
                                                \begin{eq}{mass}
\mt = \mtot - i_0 \Nbart \comma \gap 
\Nbart \equiv \int_0^\infty dN \ N \ \phit(N) \comma
                                                                \end{eq}
%-----------------------------------------------------------------------
where $\mtot$ is the total monomer concentration (including
polymerized monomers), and $\Nbart$ is mean chain length.  Here
$\phit$ is normalized to unity and $N=0$ corresponds to free
initiator.  The coupling between monomer concentration and the MWD is
manifest in eq. \eqref{mass}.

Now the dynamical equations equations obeyed by $\phit$ are 
%_______________________________________________________________________
                                                \begin{eqarray}{reflecting}
&& {\partial \phit(N) \over \partial t} = - {\partial \jt(N) \over \partial N} 
\comma \drop
&& \jt(N) \equiv \vt \phit(N) - \Dt {\partial \phit(N) \over \partial N} 
\comma \ \ \ 
\jt(0) = 0 
\comma
                                                                \end{eqarray}
%-----------------------------------------------------------------------
where 
%_______________________________________________________________________
                                                \begin{eq}{vd}
\vt \equiv \kplus \mt - \vminus
\comma \gap 
\Dt \equiv (\kplus \mt + \vminus)/ 2
\period
                                                                \end{eq}
%-----------------------------------------------------------------------
Here $\kplus$ is the propagation rate constant and $\vminus$ is the
depolymerization rate.  The ``diffusion'' coefficient $\Dt$, describes
fluctuations in the rate of polymerization/depoly\-merization.
Eq. \eqref{reflecting} is identical to the diffusion dynamics of a
particle in a linear potential with time-de\-pendent slope $-\vt$ whose
time-dependent diffusion coefficient is $\Dt$ (see
fig. \ref{slope_perturb}).  The novelty here is that due to the
coupling between the MWD and monomer, both the slope and diffusivity
are functionals of $\phi$ itself.

The reflecting boundary condition in eq. \eqref{reflecting} at $N=0$
represents the fact that when a living chain looses all of its
monomers and becomes a free initiator molecule (\ie reaches length
$N=0$) it cannot depolymerize further and must grow again.
% as shown in fig. \ref{zerolength_special}. 
Now we restate eq. \eqref{reflecting} as
%_______________________________________________________________________
                                                \begin{eq}{phi}
{\partial \phit \over \partial t} = 
- \vt {\partial \phit \over \partial N} + \Dt {\partial^2 \phit \over
\partial N^2}
\comma \ \ \ 
\phit(0)/\phit'(0) = \Dt/\vt 
\comma
                                                                \end{eq}
%-----------------------------------------------------------------------
where $\phit' \equiv \partial \phit/\partial N$.

Now it might at first seem that despite the coupling of
eq. \eqref{mass}, the monomer subsystem would effectively uncouple
from the MWD dynamics, since one might naively expect that the monomer
dynamics do not depend on the shape of the polymer MWD but only on the
number of living chains.  This would imply simple first order kinetics
for $\mt$ resulting in exponential relaxation of the monomer
concentration, independent of the MWD dynamics.  The coupling however
arises from the existence of special chains in the MWD of zero length,
\ie free initiator molecules of which there are $\phi_t(0)$, which
unlike all other chains cannot depolymerize.
%(see fig. \ref{zerolength_special}).  
Indeed, the dynamics obeyed by $\mt$, or equivalently $\vt$, are
%-----------------------------------------------------------------------
                                                \begin{eq}{v}
{d \vt \over dt} = - {\vt \over \taufast} - {\Dt \phit(0) \over
\taufast}  \comma \ \ \ 
\taufast \equiv{f \over r \vminus} \approx {f \, \Nbarinf \over (1 - f) \vminus}
\period
                                                                \end{eq}
%-----------------------------------------------------------------------
We see clearly that the only aspect of the living MWD which monomers
see is $\phit(0)$.  Eq. \eqref{v} is derived by calculating $d \Nbart /
dt$ by multiplying eq. \eqref{reflecting} by $N$, integrating over all
$N$, using the reflecting boundary condition, and then using
eq. \eqref{mass}.  The relationship between $\taufast$ and $\Nbarinf$
follows from eq. \eqref{eqm} below.  Here we have introduced the two
basic independent dimensionless parameters of the system:
%_______________________________________________________________________
                                                \begin{eq}{rf}
f \equiv {\vminus \over \kplus \mtot} \comma \gap
r \equiv {i_0 \over \mtot} \period
                                                                \end{eq}
%-----------------------------------------------------------------------
The physical meaning of $f$ is the following: if the system were a
pure solution of unpolymerized monomer (plus solvent and initiator)
$f$ would be the ratio of backward to forward polymerization
velocities.  The value of $f$ is temperature dependent, being unity at
the polymerization temperature and smaller or larger than unity in the
polymer and non-polymer phase, respectively.  The parameter $r$,
namely the ratio of living chain to total monomer concentration, is
independent of temperature and is always much smaller than unity.  Its
smallness is related to the mean degree of polymerization being much
larger than unity in equilibrium (see eq. \eqref{eqm} below).

%%%%%%%%%%%%%%%%%%%%%%%%%%%%%%%%%%%%%%%%%%%%%%%%%%%%%%%%%%%%%%%%
%%%%%%%%%%%%%%%%%%%%%%%%%%%%%%%%%%%%%%%%%%%%%%%%%%%%%%%%%%%%%%%%

Now setting the time derivative in eq. \eqref{phi} to zero and using
eq. \eqref{mass} it is shown in Appendix A that in equilibrium the
Flory-Schultz distribution of eq. \eqref{flory} is recovered and one
has to leading order in $r$:
%_______________________________________________________________________
                                          \begin{eq}{eqm}
\Nbarinf \approx {1-f \over r} \comma \ \ 
{\minf \over \mtot} \approx f  \comma \ \
\vinf  \approx - {\vminus \over \Nbarinf} \comma \ \
\Dinf \approx \vminus \comma
%\gap (r \ll 1)
                                                                \end{eq}
%-----------------------------------------------------------------------
where $r\ll 1$ and subscript $\infty$ denotes the $t \gt \infty$ equilibrium value
for the corresponding variable.  We assumed $(1-f)/r^{1/2} \gg 1$, \ie
that the temperature is not extremely close to the polymerization
temperature.

An important feature in eq. \eqref{eqm} is that in equilibrium the
velocity, \ie the slope of the ``potential'' term in eq. \eqref{phi},
settles down to a very small negative value.  Were there no diffusion,
living chains subject to a negative velocity field would shrink to
zero length.  However due to the small magnitude of the velocity,
diffusion is strong enough to broaden the MWD up to distances of order
$\Nbarinf$, as illustrated in fig. \ref{mono_to_broad}.  We show in
the following section that any apparently small external perturbation
has an enormous effect on the velocity which, depending on the
direction of the perturbation, may become either very negative or very
positive.

%************************************************************************************
%************************************************************************************
%************************************************************************************
%************************************************************************************

\section{Response to $T$-jump;  Positive Velocity Boost}

\begin{figure}[tb]             
\centering
\includegraphics[width=8cm]{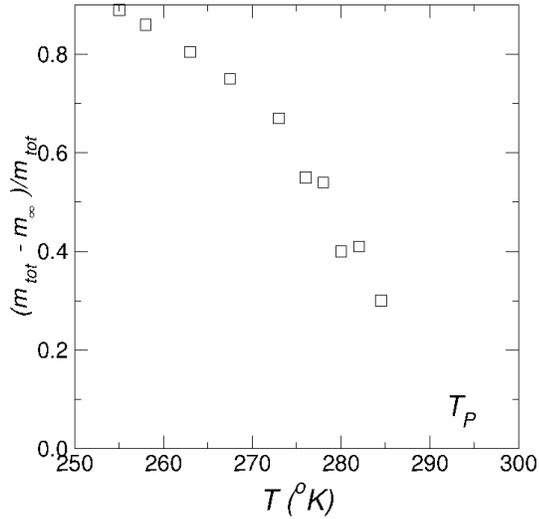}
\caption{Fraction of polymerized monomer in equilibrium as a function of
temperature for living $\alpha$-methylstyrene in tetrahydrofuran
solvent, initiated by sodium napthalene.  Data from ref.
\cite{sarkar:greer:living_4}.  Ratio of mole fraction of total
monomer, $\mtot$, to initiator is 0.0044.  Mole fraction of total
monomer is 0.1538.  The polymerization temperature $T_{\rm p}$ is near
$295^\circ K$. 
}
\label{conversion}
\end{figure}

Perhaps the simplest way to perturb an equilibrium living polymer
system is by a small temperature change ($T$-jump).  Taking as an
example $\alpha$-methylstyrene, the data of fig. \ref{conversion} show
equilibrium fraction of polymerized monomer as a function of
temperature; evidently, changes in $T$ lead to different values of the
equilibrium monomer concentration $\minf$ and therefore of $\Nbarinf$
as well.  In this section we consider relaxation dynamics after an
equilibrated system is subjected at $t=0$ to a small temperature
change $\delta T$ such that the system will eventually reach new
equilibrium values $\minf$ and $\Nbarinf$.  That is, we follow the
dynamics of the transition from an old equilibrium towards a slightly
different new equilibrium state.

The magnitude of the perturbation is measured by the small parameter
%_______________________________________________________________________
                                                \begin{eq}{edef}
\epsilon \equiv 
{\delta m_0 \over \minf}
\comma 
                                                                \end{eq}
%-----------------------------------------------------------------------
where we define $\delta \mt \equiv \mt - \minf$ and similarly for
other quantities.  All equilibrium values refer to the destination
($t=\infty$) equilibrium values.  Thus $\delta m_0$ is the initial
($t=0$) deviation from the final equilibrium.

%%%%%%%%%%%%%%%%%%%%%%%%%%%%%%%%%%%%%%%%%%%%%%%%%%%%%%%%%%%%%%%%
\begin{figure}[tb]             
\centering
\includegraphics[width=8cm]{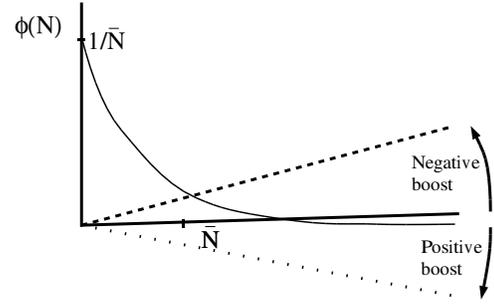}
%\epsfxsize=10cm
%\centerline{\epsffile{figures/slope_perturb.eps}}
%{  Figure \ref{zerolength_special}}
\caption[]
{  Living polymer dynamics are equivalent to diffusion
in a potential with slope $-\vt$.  In equilibrium the slope 
reaches a very small value of order $1/\Nbarinf$ (solid line)
resulting in a broad MWD.  Under a small $T$-jump the slope is 
strongly perturbed and depending
on the direction of the jump becomes either 
positive (dashed line) or negative (dotted line). 
}
\label{slope_perturb}
\end{figure}
%%%%%%%%%%%%%%%%%%%%%%%%%%%%%%%%%%%%%%%%%%%%%%%%%%%%%%%%%%%%%%%%

The value of $\epsilon$ is related to the magnitude of the 
$T$-jump.  For example in the case of $\alpha$-methylstyrene,
assuming that the system is initially below the polymerization
temperature $T_p$ in the region of fig. \ref{conversion} between
260$^\circ$K and 280$^\circ$K, one has using eq. \eqref{edef} 
%_______________________________________________________________________
                                                \begin{eq}{dT}
\epsilon \approx - {\delta T \over 50 ^\circ{\rm K}}
\gap
\mbox{($\alpha$-methylstyrene)} \period
                                                                \end{eq}
%-----------------------------------------------------------------------
Thus for $\alpha$-methylstyrene a temperature increase, $\delta T >
0$, results in a negative $\delta m_0$, \ie a reduced initial monomer
concentration relative to the destination equilibrium value.

\begin{figure}[tb]             
\centering
\includegraphics[width=9cm]{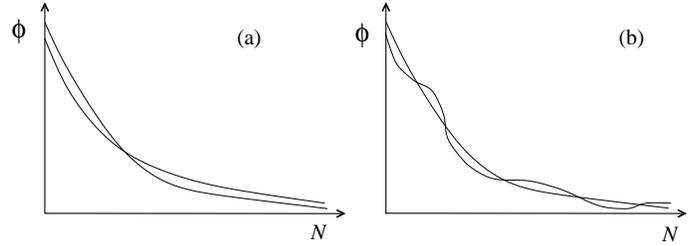}
\caption[Sketch of Initial and Final MWDs]
{(a) Sketch of the initial and final MWDs during a $T$-jump experiment.
Both distributions are exponential. Though very close to one another, 
in the course of relaxation the time-dependent MWD becomes
very different to the final exponential MWD. (b) General
perturbation (non-exponential initial MWD).
Most $T$-jump results derived here apply to this general case also.
}
\label{mwd_perturb}
\end{figure}

Now the $t=0$ relative perturbation of the MWD  is 
%_______________________________________________________________________
                                                \begin{eq}{jump}
{\delta \phi_0(N) \over  \phiinf(N) }
 \approx {\partial \phiinf(N) \over \partial
\Nbarinf}{\delta \Nbar_0 \over \phiinf (N) }
= {\epsilon \over \theta} \paren{1 - {N \over \Nbarinf}}
\comma 
\theta \equiv {1-f \over f}
\period
                                                                \end{eq}
%-----------------------------------------------------------------------
This is of order $\epsilon$ and smaller than unity for all $N$ (apart
from the unimportant values $N \gg \Nbarinf$ where the MWD is
exponentially small) as shown in fig. \ref{mwd_perturb}(a).

Because velocity, monomer concentration and mean chain length are all
linearly related (see eqs. \eqref{mass},\eqref{eqm} and \eqref{vd})
changes in these quantities are simply related as
%_______________________________________________________________________
                                                \begin{eq}{snow}
{\delta m_t \over \minf} \,
\approx - \theta \, {\delta \Nbart \over \Nbarinf} \,
\approx \,
2 {\delta \Dt \over \Dinf} \,
\approx \,
{\delta \vt \over \vminus} 
\period
                                                                \end{eq}
%-----------------------------------------------------------------------
These relations are true for all times. They allow us to follow the
dynamics of velocity alone.  Once $\vt$ is known, $\Nbart$, $\mt$,
$\Dt$ are determined. 

From eqs. \eqref{snow} and \eqref{edef} the initial relative
changes in $v, D$ are:
%_______________________________________________________________________
                                                \begin{eq}{beware}
{ \delta v_0  \over \vinf} \approx - \epsilon \, \Nbarinf  \comma
\gap
{\delta D_0 \over \Dinf} \approx  {\epsilon \over 2} \period
                                                                \end{eq}
%-----------------------------------------------------------------------
The important point is that since $\Nbarinf \gg 1$, the relative
change in $v$ is much larger than $\epsilon$ : the velocity is highly
perturbed, as shown in fig. \ref{slope_perturb}.  Depending on the
sign of $\epsilon$, the velocity may remain negative as in
equilibrium, or it may change sign. Since the relative perturbation in
$D$ is by contrast small, we see that the delicate velocity-diffusion
balance which sustained equilibrium is now destroyed.

%%%%%%%%%%%%%%%%%%%%%%%%%%%%%%%%%%%%%%%%%%%%%%%%%%%%%%%%%%%%%%%%
%%%%%%%%%%%%%%%%%%%%%%%%%%%%%%%%%%%%%%%%%%%%%%%%%%%%%%%%%%%%%%%%

We will show below that relaxation to equilibrium now occurs in three
stages.  During the first, dominated by translational motion, the MWD
is boosted far from its equilibrium shape.  The next stage involves
diffusive restoration of the region of the MWD which suffered maximum
distortion during the first stage. The third and final episode entails
a very slow diffusion-dominated recovery of the global MWD.  For the
remainder of this section we treat the case of a positive initial
velocity boost, where chains initially grow ($\delta T <0$ for
$\alpha$-methylstyrene).  Negative velocity boosts, where chains
initially shrink, are considered in section 5.

\subsection{Coherent Chain Growth: Hole Development}

\begin{figure*}[tb]             
\centering
\includegraphics[width=15cm]{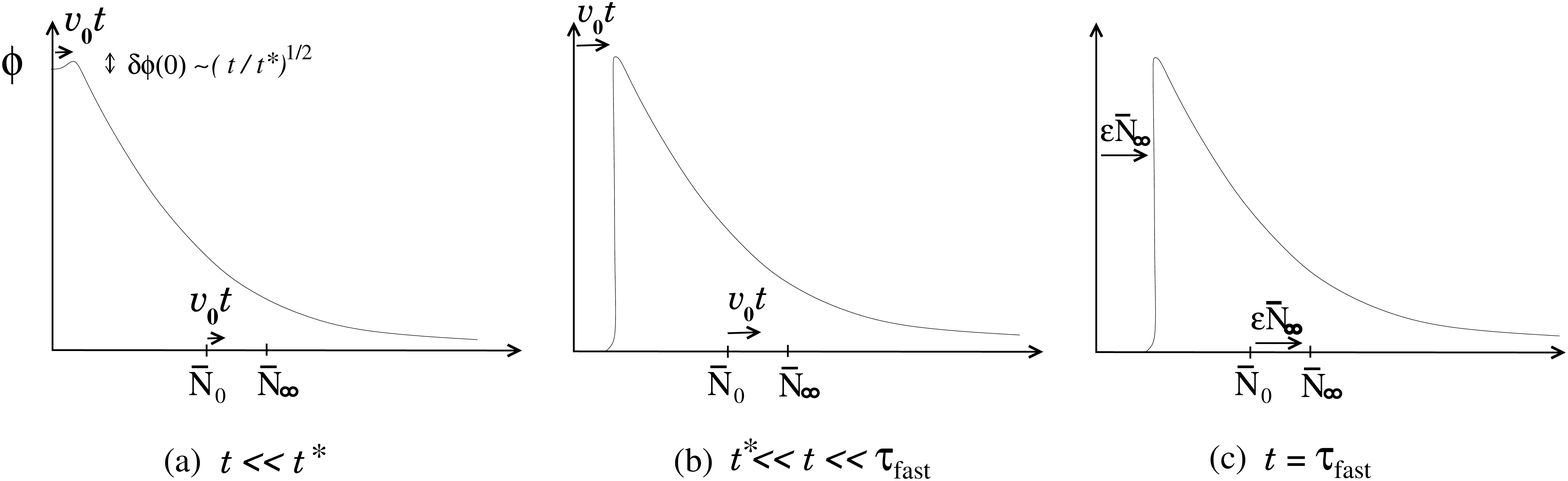}
\caption{Initial response of the MWD after a positive boost, case of strong
perturbation ($\tstar \ll \taufast$). 
(a) For small times, chain end
``diffusion'' smooths out the hole which would otherwise form at
small $N$ due to chain growth.
(b) At larger times coherent chain growth beats diffusion and a hole
develops in the MWD.
(c) Coherent growth stops when mean chain length reaches a value very close to 
$\Nbarinf$, displaced by an amount $\epsilon
\Nbarinf/\theta$ from the original value $\Nbar_0$. For typical cases,
the parameter $\theta$ is of order unity.
}
\label{translation_strong}
\end{figure*}

During the first stage of the relaxation process, velocity dominates
over diffusion since $v$ has been so strongly perturbed.  Chains thus
grow coherently, consuming the excess monomer in a timescale $\taufast$
by which time translation will have halted and a highly non-linear
hole will have developed in the MWD (see
fig. \ref{translation_strong}).  Here $\taufast$, defined in
eq. \eqref{v}, is the time for the MWD to translate distance $\delta
\Nbar_0$ and reach the destination mean, \ie $\taufast \approx \delta
\Nbar_0 / \delta v_0$ (see eqs. \eqref{edef}, \eqref{beware} and using
eq. \eqref{eqm}).

To see all of this more quantitatively, consider the velocity dynamics
eq. \eqref{v}.  Initially, the velocity term on the rhs is much larger
in magnitude than the diffusion term.  We show in Appendix B that this
remains true up until a time $\tauqs$, defined below.  It follows that
for these times $\vt$ relaxes exponentially.  The same is true of
$\mt,\Nbart$ and $\Dt$ which we recall are linearly coupled to $\vt$
(see eq. \eqref{snow}).  Thus
%_______________________________________________________________________
                                                \begin{eq}{mi}
\delta \vt \approx
\epsilon\, \vminus\, e^{-t/\taufast} - \vinf
\ \ \  (t \ll \tauqs  = \epsilon \Nbarinf^{3/2}\theta^{-3/2}/ \vminus) 
\period
                                                                \end{eq}
%-----------------------------------------------------------------------
Note that for $t \gg \taufast$ there are no free initiators (at much
longer times these are restored, see below). Thus for these
intermediate times all chains are identical as far as on and off
monomer kinetics are concerned, and the net rate of monomer-polymer
mass exchange can only vanish if $\vt$ vanishes. For this reason $\vt$
decays to zero, though much later it will recover its small negative
equilibrium value. The $\phit$ dynamics, eq. \eqref{phi}, thus
simplify to
%_______________________________________________________________________
                                                \begin{eq}{eu}
{\partial \phit \over \partial t} \approx 
-  \epsilon\, \vminus e^{-t/\taufast} 
 {\partial \phit \over \partial N} +
 \Dinf {\partial^2 \phit \over \partial N^2} 
\ \ \ \ \ (t \ll \tauqs)
                                                                \end{eq}
%-----------------------------------------------------------------------
where the diffusion coefficient may be approximated by $\Dinf$ since
the contribution of $\delta \Dt$ is negligible (see Appendix
B).  This linear equation, plus the time-dependent boundary condition
of eq. \eqref{phi},  has solution
%_______________________________________________________________________
                                                \begin{eq}{evolve}
\phit(N) \approx \int_0^\infty dN' \, \Gtran_t(N,N') \phi_0 (N')
\ \ \ \ \ (t \ll \tauqs)                                 
                                                \end{eq}
%-----------------------------------------------------------------------
which describes a translating and simultaneously broadening MWD, as
shown in fig. \ref{translation_strong}.  Here $\Gtran_t(N,N')$ is the
propagator of eq. \eqref{eu} whose properties are calculated in
Appendix B.

%%%%%%%%%%%%%%%%%%%%%%%%%%%%%%%%%%%%%%%%%%%%%%%%%%%%%%%%%%%%%%%%
%%%%%%%%%%%%%%%%%%%%%%%%%%%%%%%%%%%%%%%%%%%%%%%%%%%%%%%%%%%%%%%%

%%%%%%%%%%%%%%%%%%%%%%%%%%%%%%%%%%%%%%%%%%%%%%%%%%%%%%%%%%%%%%%%
%%%%%%%%%%%%%%%%%%%%%%%%%%%%%%%%%%%%%%%%%%%%%%%%%%%%%%%%%%%%%%%%

An important quantity in what follows is $\phit(0)$, \ie the
concentration of free initiators.  Using eq. \eqref{evolve} we show in
Appendix B that
%_______________________________________________________________________
                                                \begin{eqarray}{approximate}
&& {\phit (0) / \phiinf(0)}  \approx \drop
&& \casesbracketsshortii{1 - C \, (t/\tstar)^{1/2}}{t \ll \tstar}
                {F\, (\tstar/t)^{3/2} e^{-t/\tstar}}
                                                 {\tstar \ll t \ll \taufast}
\ \ \tstar \equiv {4 \Dinf \over  v_0^2}
\comma
                                                                \end{eqarray}
%-----------------------------------------------------------------------
where $C=4/\pi^{1/2}$ and $F=\pi^{-1/2}$. Eq. \eqref{approximate} has
a clear physical meaning: as the MWD translates, the position of its
peak at $N=0$ after time $t$ has shifted to chain lengths of order
$v_0 t$.  Were diffusion absent, this would have created a depletion
region in the MWD at small chain lengths.  However diffusion smooths
out inhomogeneities on distances of size $(\Dinf t)^{1/2}$.  Hence for
times shorter than the crossover time $\tstar$ for which $v_0 \tstar
\approx (\Dinf \tstar)^{1/2}$, diffusion has enough time to fill the
translationally induced hole and the relative deviation from
equilibrium is small.  For times longer than $\tstar$ however, the MWD
translates a distance away from the origin much larger than what
diffusion could have homogenized, a hole develops in the MWD, and the
concentration of zero length chains becomes exponentially small.

In eq. \eqref{approximate} we assumed that the perturbation is large
enough such that $\tstar \ll \taufast$.  The special case of extremely
small perturbations where this is no longer true is discussed
separately in section 5.

\subsection{Diffusive Length Relaxation: Hole Filling}

For times $t \gg \taufast$ the fast variables monomer and mean
chain length have relaxed very close to their equilibrium values.  We
have seen that velocity becomes exponentially small. Meanwhile (see
below) the number of free initiators $\phit(0)$ is gradually
recovering. At a certain moment, therefore, the 2 terms on rhs of
eq. \eqref{v} become equal to one another and we show in Appendix B
that for all later times the velocity time derivative is much smaller.
Hence the velocity dynamics are now fundamentally changed.  The new
regime is one of quasi-static evolution, enslaved to the dynamics of
$\phit(0)$:
%_______________________________________________________________________
\begin{eq}{all} 
\delta \vt
\approx \
%\casesbracketsshortii {\epsilon \, e^{-t/\taufast} + r/(1-f)}
%{t \ll \tauqs}
{- \vminus \delta \phit(0)} 
\gap (t \gg \tauqs) \period
%\gap \tauqs = \Nbarinf^{3/2}/ \vminus 
\end{eq}
%-----------------------------------------------------------------------
Physically, this reflects the fact that the the only aspect of the MWD
shape seen by monomers is the amount of special non-depolymerizing
zero-length chains.  We see that as the MWD slowly rearranges itself,
so the fast variables $\vt,\mt$ and $\Nbart$ variables respond
quasi-statically.

Notice that the quasi-static regime does not onset after $\taufast$ but
rather after $\tauqs$.  The reason is that during the initial boost
the term $\Dt \phi_t(0)$ in eq. \eqref{v} decays exponentially on a
timescale $\tstar$; its magnitude therefore at $\taufast$ is much
smaller than $\vt$ which decays on a timescale $\taufast$ (see
eqs. \eqref{mi} and \eqref{approximate}).  There is thus a
cross-over period before the quasi-static balance between the two
terms is established.

Now let us examine the MWD dynamics.  Since $\vt$ becomes
exponentially small after the MWD stops translating, we have
%_______________________________________________________________________
                                              \begin{eq}{simple} 
{\partial \phit \over \partial t} 
\approx 
\Dinf {\partial^2 \phit \over \partial N^2} 
\gap (\taufast \ll t \ll \taufill) 
                                               \end{eq}
%-----------------------------------------------------------------------
with reflecting boundary conditions at the origin.  Here once again
the contribution of $\delta \Dt$ is negligible (see eqs. \eqref{snow},
\eqref{all}, and \eqref{longphi}) and the corresponding term has been
neglected in eq. \eqref{simple}.  Here
%_______________________________________________________________________
                                                 \begin{eq}{tautwo}
\taufill \equiv \epsilon^2 \, \theta^{-2} \, \tauslow  \comma \gap
\tauslow \equiv {\Nbarinf^2 \over 4 \Dinf}
\period
                                                                 \end{eq}
%-----------------------------------------------------------------------
The timescale $\tauslow$ is the longest relaxation time of the system,
the time for the slow global shape characteristics of the MWD to
relax. It equals the diffusion time for the MWD width $\Nbarinf$.
Meanwhile $\taufill$ is the diffusion time for the hole width,
$\epsilon \Nbarinf/\theta$. It is shown in Appendix B that
eq. \eqref{simple} has solution
%_______________________________________________________________________
                                                 \begin{eq}{longphi}
{\phit (0) \over \phiinf(0)} \approx 
%\casesbracketsshortii
H\, \paren{t \over \taufill}^{1/2} e^{-\taufill/t}
        \ \ \ \   (\taufast \ll t \ll \taufill)
%{1-J \, \paren{\taufill/t}^{1/2}}
%                                {\taufill \ll t \ll \tauslow}
                                                                 \end{eq}
%-----------------------------------------------------------------------
where $H=\pi^{-1/2}$.  Thus for times shorter than $\taufill$ the
concentration of zero length chains remains exponentially small while
for longer times the hole fills (see fig. \ref{hole_fillup}) and
$\phit(0)$ recovers its equilibrium value.

\subsection{Linearized Dynamics at Long Times}

For times beyond $\taufill$ the hole-filling is complete and the MWD's
non-linear feature has disappeared. Thus, finally, a truly linear
regime onsets: relative deviations of all variables from equilibrium
are less than unity and perturbation theory can now be applied. This
is done in Appendix C where we find
%_______________________________________________________________________
                                                 \begin{eqarray}{expo}
&& {\delta \vt \over \vminus} 
\approx
- \delta \phit (0)  
\ \approx \drop 
&& \casesbracketsshortii
{J \, \phiinf(0) \, (\taufill /t)^{1/2}}        
                                        {\taufill \ll t \ll \tauslow}
{M \, \epsilon \phiinf(0) (t / \tauslow)^{-3/2} e^{-t/(16 \tauslow)}}
                                        {t \gg \tauslow}
\drop
                                                                 \end{eqarray}
%-----------------------------------------------------------------------
where $J, M$ are positive constants of order unity.  Thus $\delta
\phit(0)$, and all of the fast variables which remain quasi-statically
enslaved to $\phit(0)$, relax exponentially on a time\-scale $\tauslow$
with a power law prefactor.

%************************************************************************************
%************************************************************************************
%************************************************************************************
%************************************************************************************

\section{The Second Moment of the MWD Relaxes More Slowly than the First}

A picture has emerged of fast mass-related variables which relax in
$\taufast$ and slow MWD shape properties which relax in the much
longer time $\tauslow$. The MWD's first moment defines the total mass
in the polymer system: this is a fast variable and was considered in
section 3. The simplest slow shape property is the dispersion
$\Deltat$,
%_______________________________________________________________________
                                                \begin{eq}{second}
\Deltat \equiv \Nbartsquared - \Nbart ^2 \comma 
\gap
\Nbartsquared \equiv \int_0^\infty dN \, N^2 \phit(N) \comma
                                                                \end{eq}
%-----------------------------------------------------------------------
which is closely related to the second moment. In this section we
consider the very different and slow relaxation of this quantity.

Substituting eq. \eqref{jump} in eq. \eqref{second} one finds that
following a $T$-jump the initial relative perturbation in the
dispersion, $\delta \Delta_0$, is 
%_______________________________________________________________________
                                                \begin{eq}{flagella}
{\delta \Delta_0 \over \Deltainf}
= - 2 \, { \epsilon \over  \theta } \comma
\gap
\Deltainf = \Nbarinf^2 \comma
                                                                \end{eq}
%-----------------------------------------------------------------------
where the equilibrium value, $\Deltainf$, is found using
eq. \eqref{flory}.  Now the dynamics of the dispersion can be derived
by evaluating the time derivatives of $\Nbartsquared, \Nbart$ in
eq. \eqref{second}.  These can in turn be calculated by multiplying
eq. \eqref{phi} by $N^2$ and $N$, respectively, integrating over all
$N$ and using the boundary condition of eq. \eqref{reflecting}.  One
has
%_______________________________________________________________________
                                                \begin{eqarray}{nature}
\inverse{\Deltainf} {d \delta \Deltat \over dt} 
\, = \,
\inverse{2 \tauslow} {\Dt \over \Dinf} 
\square{1 - {\Nbart \over \Nbarinf} {\phit(0) \over \phiinf(0)}}
\, \approx \, \drop
- \inverse{2 \tauslow} {\delta \phit(0) \over \phiinf(0)} \period
                                                                \end{eqarray}
%-----------------------------------------------------------------------
Here we replaced $\Dt$ and $\Nbart$ with their $t=\infty$ values. This
is correct to leading order since the relative perturbations in these
quantities are much smaller than that of $\phit(0)$ which undergoes
much larger changes (see section 3).

Now integrating eq. \eqref{nature} using eqs. \eqref{approximate},
\eqref{longphi} and \eqref{expo} one finds that up to the time $\taufill$
the dispersion has changed very little relative to its value just
after the perturbation at $t=0$:
%_______________________________________________________________________
                                                \begin{eq}{brown}
{\Deltat- \Delta_0 \over \Deltainf} 
\, \approx \,
\left\{
\begin{array}{llll}
{(C/3) (t/\tauslow)(t/\tstar)^{1/2}}    
                       \            & {(t \ll \tstar)}    &\blank  
{t/(2\tauslow)}
                       \            & {(\tstar \ll t \ll \taufill)}  &\blank
\end{array}%
\right.
                                                                \end{eq}
%-----------------------------------------------------------------------
This is in stark contrast to the first moment which had already
relaxed by the much shorter time $\taufast$.  The dispersion's
relaxation process does not properly get going until times of order
$\tauslow$:
%_______________________________________________________________________
                                                \begin{eq}{silver}
{\delta \Deltat \over \Deltainf}
\, \approx \,
\left\{
\begin{array}{llll}
{ \delta \Delta_0 / \Deltainf + J \epsilon \theta^{-1} (t/\tauslow)^{1/2}} 
                       \            & {(\taufill \ll t \ll \tauslow)}  &\blank
{ \alpha\, \epsilon \, (t/\tauslow)^{-5/2} e^{-t/(16 \tauslow)}} 
                        \          &  {(t \gg \tauslow)}   &\blank
\end{array}%
\right.
                                                                \end{eq}
%-----------------------------------------------------------------------
where $\alpha=8 M /39$. The relative perturbation in the dispersion
remains of order $\epsilon$ for $t \ll \tauslow$ and subsequently
relaxes exponentially.

%%%%%%%%%%%%%%%%%%%%%%%%%%%%%%%%%%%%%%%%%%%%%%%%%%%%%%%%%%%%%%%%
\begin{figure}[tb]             
\centering
\includegraphics[width=8.5cm]{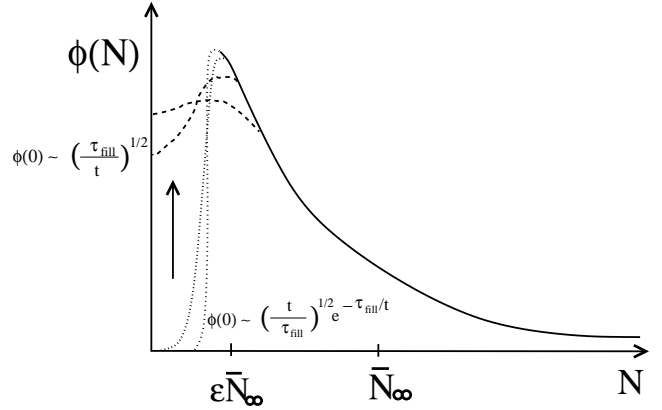}
\caption{Schematic of MWD during hole-filling episode.  
Dotted (dashed) lines show MWD profile for 
times $\taufast \ll t \ll \taufill$ ($\taufill \ll t\ll \tauslow$).
}
\label{hole_fillup}
\end{figure}
%%%%%%%%%%%%%%%%%%%%%%%%%%%%%%%%%%%%%%%%%%%%%%%%%%%%%%%%%%%%%%%%

%************************************************************************************
%************************************************************************************
%************************************************************************************
%************************************************************************************

\section{Response to $T$-jump; Negative Velocity Boost}

%%%%%%%%%%%%%%%%%%%%%%%%%%%%%%%%%%%%%%%%%%%%%%%%%%%%%%%%%%%%%%%%
\begin{figure*}[tb]             
\centering
\includegraphics[width=15cm]{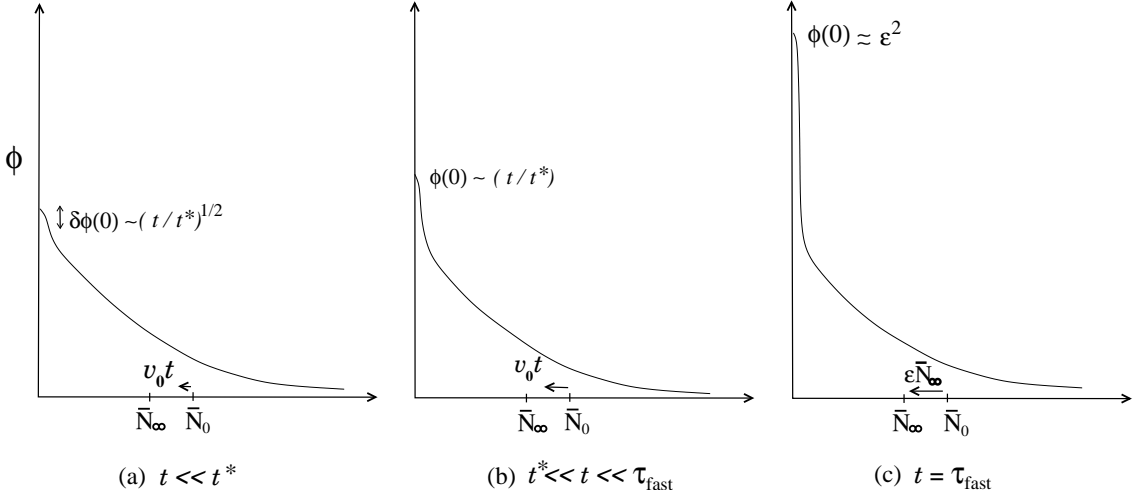}
\caption
{Initial response of MWD after a negative boost (strong
perturbation, $\tstar \ll \taufast$). 
(a) Chains shrink coherently and MWD moves from the initial mean
$\Nbar_0$ towards $\Nbarinf$.  For small times, diffusion smooths out
excess peak at small $N$.
(b) Beyond $\tstar$ diffusion is too slow to counteract coherent chain
shrinkage and
short chains accumulate at the origin.
(c) Coherent shrinkage halts when mean chain length approaches
$\Nbarinf$. 
}
\label{trans_strong_peak}
\end{figure*}
%%%%%%%%%%%%%%%%%%%%%%%%%%%%%%%%%%%%%%%%%%%%%%%%%%%%%%%%%%%%%%%%

In this section we study the case opposite to section 3, when
$\epsilon <0$ and the $T$-jump induces a higher equilibrium monomer
concentration and a smaller $\Nbarinf$ (see eq. \eqref{edef}).  Such a
perturbation in the example of $\alpha$-methylstyrene would be
produced by an increase in temperature as one may see in
fig. \ref{conversion}. In this case the sign of the initial perturbed
velocity (eq. \eqref{beware}) is negative (see
fig. \ref{slope_perturb}).  Thus initially the MWD is not boosted
towards larger molecular weights as in section 3, but is instead
boosted in the opposite direction, towards {\em smaller} $N$.
Therefore, instead of depletion, an excess of small length chains and
free initiators is produced, creating instead of a hole, a sharp peak
in the MWD near $N=0$ as shown in fig. \ref{trans_strong_peak}.

Despite the differences between the two cases, many of the results of
section 3 remain unchanged.  In fact all equations of section 3 up to
eq. \eqref{evolve} remain unchanged, the only difference being that
$\epsilon$ is negative.  Thus eq. \eqref{evolve}
now describes a MWD coherently shrinking and simultaneously broadening
with diffusion coefficient $\Dinf$.  But since free initiators cannot
depolymerize, excess polymer must build up near $N=0$.  Indeed,
starting from eq. \eqref{evolve} it is shown in Appendix D that
%_______________________________________________________________________
                                                 \begin{eqarray}{peak}
{\phit (0) \over \phiinf(0)} \approx 
\casesbracketsshortii{1 + C' \, (t/\tstar)^{1/2}}{t \ll \tstar}
                 {F'\, (t/\tstar)}
                                                  {\tstar \ll t \ll \taufast}
                                                                 \end{eqarray}
%-----------------------------------------------------------------------
where $C'=4/\pi^{1/2}$ and $F' = 4$.  Eq. \eqref{peak} has a similar
interpretation to eq. \eqref{approximate}.  For $t \ll \tstar$
diffusion is fast enough to smooth out the excess polymer accumulated
by the negative velocity at the origin and thus $\phit(0)$ remains
close to its initial value (see fig. \ref{trans_strong_peak} (a)).
For $t \gg \tstar$ however the flow towards the origin is faster than
what diffusion can smear out and a peak forms whose height increases
with time, as shown in fig. \ref{trans_strong_peak}(b).  (Once again
it is assumed that the perturbation is not so small that $\tstar \ll
\taufast$; the opposite case is examined in section 5).

After the completion of the first stage at times of order $\taufast$,
similarly to section 3, the fast $\mt, \Nbart$ variables have
essentially relaxed and start to respond quasi-statically to the slow
diffusion-driven shape changes of the MWD.  In fact we show in
Appendix D that eqs. \eqref{mi} and \eqref{all} still apply, but now
with a different crossover timescale, $\tauqs = \taufast \ln
\Nbarinf$.  This difference arises because unlike the positive boost 
case $\phit(0)$ does not become exponentially small and so needs less
time to catch up with the $\vt$ term in the velocity dynamics,
eq. \eqref{v}.

Similarly to the positive boost, eq. \eqref{simple} still applies for
$t < \taufill$.  By this time the accumulation at the origin is able
to diffuse to distances of the same size as the region from which it
was transferred from during the first stage.  Hence $\phit(0)$ becomes
of order its initial value, and thereafter perturbation theory is
valid.  In Appendix D we show that
%_______________________________________________________________________
                                                 \begin{eqarray}{phispread}
&& {\phit (0) / \phiinf(0)} \approx \drop
&& \casesbracketsshortiii
 {K' (\taufill/t)^{1/2}}
                                 {\taufast \ll t \ll \taufill}
 {1+J' (\taufill/t)^{1/2}}
                                 {\taufill \ll t \ll \tauslow}
 {1 + M' \epsilon \, (t/ \tauslow)^{-3/2} e^{-t/(16 \tauslow)}}
                                 {t \gg \tauslow} \drop
                                                \end{eqarray}
%-----------------------------------------------------------------------
where $K',J',$ and $M'$ are positive constants of order unity.  Note
there is apparently no smooth cross-over at $t=\taufast$ between the
forms of eqs. \eqref{peak} and \eqref{phispread}.  This is because
there is in fact an extra episode at times of order $\taufast$ during
which the very thin peak spreads from width $(\Dinf \tstar)^{1/2}$ to
a width $(\Dinf \taufast)^{1/2}$. Note also that the time behavior for
$t \gg \taufill$ has been derived using the linearized dynamics of
Appendix C.

Finally, to obtain the dispersion dynamics we use eqs. \eqref{peak}
and \eqref{phispread} in eq. \eqref{nature} to obtain
%_______________________________________________________________________
                                                \begin{eqarray}{gold}
&& {\delta \Deltat / \Deltainf} - 
{\delta \Delta_0 / \Deltainf}
\, \approx \drop
&& \left\{
\begin{array}{ll}
{-(C'/3) (t/\tauslow)(t/\tstar)^{1/2}}  
                       \hspace{0.5cm}              {(t \ll \tstar)}    &\blank  
{-F'/4 (t/\tauslow)(t/\tstar)}  
                       \hspace{1.2cm}     {(\tstar \ll t \ll \taufast)}    &\blank  
{-K' \epsilon \theta^{-1}  (t/ \tauslow)^{1/2}}
                       \hspace{1.3cm}        {(\taufast \ll t \ll \taufill)}  &\blank
{-J' \epsilon \theta^{-1} (t/\tauslow)^{1/2}} 
                       \hspace{1.4cm}       {(\taufill \ll t \ll \tauslow)}  &\blank
{- \delta \Delta_0 / \Deltainf - (8 M' /39) \, \epsilon \, (\tauslow/t)^{5/2} e^{-t/(16 \tauslow)}} 
                                  & \blank
                        \hspace{4.4cm} {(t \gg \tauslow)} & 
\end{array}%
\right. 
                                                                \end{eqarray}
%_______________________________________________________________________
which is very similar to eqs. \eqref{brown} and \eqref{silver} for the
positive boost case.  Again the dispersion relaxes on a timescale
$\tauslow$, much longer then the relaxation time of $\Nbart$.

%************************************************************************************
%************************************************************************************
%************************************************************************************
%************************************************************************************

\section{Very Small $T$-jumps: $\epsilon$ less than its critical value
$\epsilonc$}

Hitherto we implicitly assumed sufficiently strong velocity boosts
that the timescale $\tstar$ after which translational motion dominates
diffusion and a hole or peak begins to form is shorter than the time
for full development of the hole or peak, namely $\taufast$.  Now the
ratio of these timescales is
%_______________________________________________________________________
                                                 \begin{eq}{ecrit}
 {\tstar \over \taufast} \approx \paren{\epsilonc \over \epsilon}^2
 \comma \gap
 \epsilonc \equiv {\theta^{1/2}\over\Nbarinf^{1/2}} \comma
                                                                 \end{eq}
%-----------------------------------------------------------------------
which defines a critical value of the perturbation parameter,
$\epsilonc$. Clearly, the system's response must be different for
$\epsilon$ values below this rather small threshold. We consider such
very small perturbations in this section.  Although a full hole or
peak does not have time to develop, we will find that the response is
still strong and non-linear.

Since such weak perturbations do not involve the creation of a
complete hole or a peak in the MWD, one expects that the timescales
$\tstar$ and $\taufill$ associated with the their creation and
destruction lose their physical meaning as cross-over timescales.
This is shown in Appendix E where, using similar arguments to the ones
of sections 3 and 5, it is proved that weak perturbations are
identical to stronger perturbations except that the regime up to
$\tstar$ is truncated at $\taufast$ and the following regime (which
for stronger perturbations lasted till $\taufill$) deleted.  In Appendix
E it is shown that for both positive and negative boosts, one has for
the fast variables
%_______________________________________________________________________
                                                 \begin{eq}{weak-all}
 {\delta \vt \over \vminus}  
 \ \approx \ 
 \casesbracketsshortii
 {\epsilon \, e^{-t/\taufast}}             {t \ll \taufast}
 {- \delta \phit(0)}            {\taufast \ll t \ll \tauslow}
                                                                 \end{eq}
%-----------------------------------------------------------------------
Note that that the timescale $\tauqs$ does not appear since its magnitude lies
between those of $\tstar$ and $\taufill$. 

%%%%%%%%%%%%%%%%%%%%%%%%%%%%%%%%%%%%%%%%%%%%%%%%%%%%%%%%%%%%%%%%
\begin{figure}[tb]             
\centering
\includegraphics[width=8.4cm]{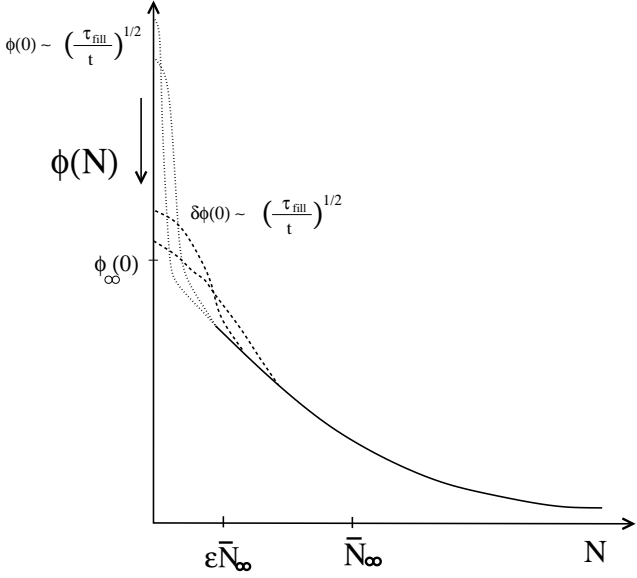}
\caption[Spreading of Small Chain Length Peak: Strong Perturbation, Negative
Velocity Boost] 
{Schematic of  MWD development during
$\taufast \ll t\ll \tauslow$
when the small chain peak spreads diffusively.
Dotted (dashed) lines correspond to
times $\taufast \ll t \ll \taufill$ ($\taufill \ll
t \ll \tauslow$).
}
\label{peak_spread}
\end{figure}
%%%%%%%%%%%%%%%%%%%%%%%%%%%%%%%%%%%%%%%%%%%%%%%%%%%%%%%%%%%%%%%%

The evolution of the MWD (see figs. \ref{translation_weak},
\ref{trans_weak_peak}) is similarly shown in Appendix E to lead to the
following solution for $\phit(0)$ in the case of a positive initial
velocity:
%_______________________________________________________________________
                                                 \begin{eq}{pon}
 {\phit (0) \over \phiinf(0)} \approx 
 \casesbracketsshortii
 {1 - C \, (t/\tstar)^{1/2}}
                                 {t \ll \taufast}
 {1- Q \paren{\taufill/t}^{1/2}}
                                 {\taufast \ll t \ll \tauslow}
 \
 (v_0 > 0) 
                                                                 \end{eq}
%-----------------------------------------------------------------------
while for a negative initial velocity one has instead
%_______________________________________________________________________
                                                 \begin{eq}{pin}
 {\phit (0) \over \phiinf(0)} \approx 
 \casesbracketsshortii
 {1 + C' (t/\tstar)^{1/2}}
                                 {t \ll \taufast}
 {1 + Q' (\taufill/t)^{1/2}}
                                 {\taufast \ll t \ll \tauslow}
 \
 (v_0 < 0)
                                                                 \end{eq}
%-----------------------------------------------------------------------
where $Q,Q'$ are positive constants.  Finally, it is straightforward
to show that the relaxation for for $t \gg \tauslow$ is exponential as
for the stronger perturbations, eq. \eqref{expo}.

%IIIIIIIIIIIIIIIIIIIIIIIIIIIIIIIIIIIIIIIIIIIIIIIIIIIIIIIIIIIIIIIIIIII
\ignore{
 with an extra negative sign in
front of the last term for $\epsilon < 0$.
} % end \ignore{IIIIIIIIIIIIIIIIIIIIIIIIIIIIIIIIIIIIIIIIIIIIIIIIIIIII

The most important result of this section is that even though the
$\epsilon < \epsilonc$ case is ``weak'' as compared to sections 3 and
4, the system's response remains large and nonlinear.  One sees from
eqs. \eqref{pon} and \eqref{pin} that at $\taufast$ when the deviation
of $\phit(0)$ from equilibrium is largest, the relative deviation of
$\phit(0)$ from equilibrium is much larger than $\epsilon$.  Therefore
even in the weak case a hole or peak does form in the MWD but of a
smaller amplitude as compared to the stronger perturbation case (see
figs. \ref{translation_weak}, \ref{trans_weak_peak}).

Now the dispersion dynamics may be similarly evaluated using
eqs. \eqref{pon} and \eqref{pin} in eq. \eqref{nature}.  One finds
that $\Deltat$ relaxes once again on a timescale $\tauslow$.  For the
positive initial velocity case one finds 
%_______________________________________________________________________
                                                \begin{eqarray}{bronze}
&& {\delta \Deltat / \Deltainf} - 
{\delta \Delta_0 / \Deltainf}
\, \approx \drop
&& \left\{
\begin{array}{ll}
{(C/3) (t/\tauslow)(t/\tstar)^{1/2}}    
                       \hspace{0.5cm}   {(t \ll \taufast)}    &\blank  
{Q \epsilon \theta^{-1}  (t/\tauslow)^{1/2}} 
                       \hspace{1.3cm} {(\taufast \ll t \ll \tauslow)}  &\blank
{- \delta \Delta_0 / \Deltainf + (8 M /39) \, \epsilon \, (\tauslow/t)^{5/2} e^{-t/(16 \tauslow)}} 
                                  &\blank
\hspace{4cm} {(t \gg \tauslow)}   &\blank
\end{array}%
\right. 
                                                                \end{eqarray}
%-----------------------------------------------------------------------
while for a negative initial boost
%_______________________________________________________________________
                                                \begin{eqarray}{wood}
&& {\delta \Deltat / \Deltainf} - 
{\delta \Delta_0 / \Deltainf}
\, \approx \drop
&& \left\{
\begin{array}{ll}
{-(C'/3) (t/\tauslow)(t/\tstar)^{1/2}}  
                       \hspace{0.5cm} {(t \ll \taufast)}    &\blank  
{-Q' \epsilon \theta^{-1} (t/\tauslow)^{1/2}} 
                       \hspace{1.3cm} {(\taufill \ll t \ll \tauslow)}  &\blank
{- \delta \Delta_0 / \Deltainf - (8 M' /39) \, \epsilon \, (\tauslow/t)^{5/2} e^{-t/(16 \tauslow)}} 
                        &\blank
\hspace{4.4cm}  {(t \gg \tauslow)}   &\blank
\end{array}%
\right. 
                                                                \end{eqarray}
%_______________________________________________________________________

%************************************************************************************
%************************************************************************************
%************************************************************************************
%************************************************************************************

\section{General Perturbations}

We have so far considered a specific type of perturbation, namely
$T$-jumps.  In this section we consider general perturbations.  We
will see that almost all of the phenomenology remains unchanged.

Let us start with another simple type of perturbation: addition of a
very small amount of monomer to an equilibrated system.  (This would
require special care to avoid introducing impurities which in current
experimental methods are destroyed by the initiators before
polymerization \cite{greer:review}.)  Changing the total amount of
monomer perturbs both independent system parameters, $r$ and $f$ (see
eq. \eqref{rf}), while a $T$-jump perturbed only $f$.  Despite this,
the results of the previous sections remain identical.  The reason is
that since in both cases the initial MWD is an exponential
distribution whose mean is different from the new target equilibrium
MWD, the only variable parameterizing the perturbation is $\delta
\Nbar_0$, or equivalently $\delta m_0$.  Thus one sees from
eq. \eqref{edef} that monomer addition is simply equivalent to a
positive velocity boost induced by a $T$-jump.

Consider now the most general perturbation, \ie one generating an
initial MWD of arbitrary shape as shown in
fig. \ref{mwd_perturb}(b). (This should be compared to the $T$-jump
and $m$-jump which produce initial MWDs of exponential shape with a
mean slightly displaced from equilibrium.)  Such a general
perturbation is small provided
%_______________________________________________________________________
                                                \begin{eq}{gsmall}
{|\delta \phi_0(N)| \over \phiinf(N)} \, \lsim \, \epsilon 
                                                                \end{eq}
%-----------------------------------------------------------------------
for all $N$. As for the $T$-jump case, we have now
%_______________________________________________________________________
                                                \begin{eq}{arbitrary}
\epsilon \equiv {\delta m_0 \over \minf} 
\, \approx \,
- {\delta \Nbar_0 \over \Nbarinf} \theta
\, = \,
- {\theta\over \, \Nbarinf} \int_0^\infty dN \, N \delta \phi_0(N)  
                                                                \end{eq}
%-----------------------------------------------------------------------
Since eq. \eqref{arbitrary} is of the same form as eq. \eqref{edef},
all the arguments of section 3 leading to the large relative
perturbation in $v$ (eq. \eqref{beware}) remain unchanged.  Moreover,
in our analysis of sections 3, 4, and 5, the exact shape of the
perturbed MWD did not matter in any of the leading order results we
obtained.  The reason is that the effect of the velocity field on the
MWD is so drastic that to first order the response of any MWD close
enough to equilibrium is the same, in the sense that the MWD simply
translates, creating a hole (or peak). Thus the analysis of sections
3, 4 and 5 directly apply to the general case.

A special exception is when the first moment of the perturbation
$\delta \phi_t(0)$ is arranged to have a value much smaller than
$\epsilon \Nbarinf$.  In this case the relative amount of
monomer-polymer mass transfer is much less than $\epsilon$ and this
in turn reduces the velocity perturbation.  For example, if the first
moment were chosen to be so small that $\delta m_0/\minf \lsim
1/\Nbarinf$, then from eq. \eqref{beware} one sees that the relative
perturbation in $\vt$ is of order $\epsilon$, much less than the order
$\epsilon\Nbarinf$ produced by a $T$-jump.  In such cases, linear
response applies for all times.  However, if the first moment of
$\delta \phi_0$ is much greater than unity the response remains
nonlinear.

%%%%%%%%%%%%%%%%%%%%%%%%%%%%%%%%%%%%%%%%%%%%%%%%%%%%%%%%%%%%%%%%
\begin{figure}[tb]             
\centering
\includegraphics[width=8cm]{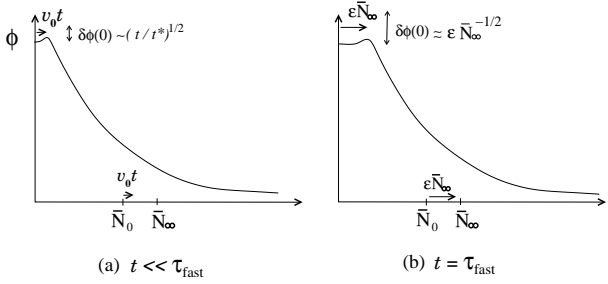}
\caption
{Weak perturbations, $\epsilon < \epsilonc$. For such small perturbations
$\tstar \gg \taufast$. 
(a) The coherent component of chain growth shifts the MWD mean
towards $\Nbarinf$.  Diffusion is dominant, smoothing depletion at
small $N$.  The net effect is a nonlinear deviation from equilibrium. 
(b) By $\taufast$ coherent growth halts, when mean chain length approaches
$\Nbarinf$.  The behavior for longer times, $t \gg \taufast$, is
essentially the same as that shown by the dashed lines in fig. \ref{hole_fillup}.
}
\label{translation_weak}
\end{figure}
%%%%%%%%%%%%%%%%%%%%%%%%%%%%%%%%%%%%%%%%%%%%%%%%%%%%%%%%%%%%%%%%

%%%%%%%%%%%%%%%%%%%%%%%%%%%%%%%%%%%%%%%%%%%%%%%%%%%%%%%%%%%%%%%%
\begin{figure}[tb]             
\centering
\includegraphics[width=8.5cm]{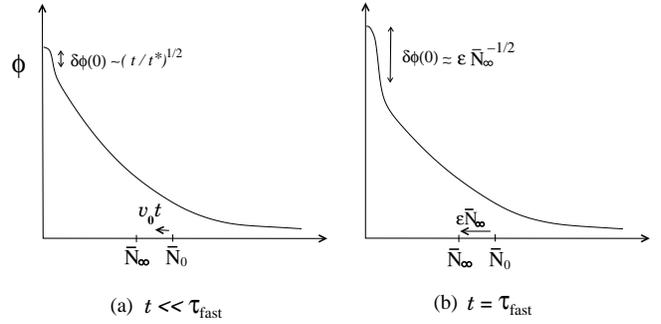}
\caption[MWD Translation: Weak Perturbation, Negative
Velocity Boost]
{As in fig. \ref{translation_weak} but now for a negative initial
velocity boost; a peak is formed instead of depletion.
Evolution for $t \gg \taufast$ as dashed lines of fig. \ref{peak_spread}.
}
\label{trans_weak_peak}
\end{figure}
%%%%%%%%%%%%%%%%%%%%%%%%%%%%%%%%%%%%%%%%%%%%%%%%%%%%%%%%%%%%%%%%

%************************************************************************************
%************************************************************************************
%************************************************************************************
%************************************************************************************

\section{Discussion}

The peculiarity of living polymers is that different moments of the
MWD relax to equilibrium on different time\-scales.  In the present work
we have shown that the first moment, \ie the mean length, $\Nbart$,
relaxes on a timescale $\taufast$.  However the shape of the MWD
measured by the second moment, or equivalently the dispersion,
$\Deltat$, relaxes after $\tauslow$.  Because the two timescales
depend on different powers of $\Nbarinf$ their ratio may be extremely
large:
%_______________________________________________________________________
                                                \begin{eq}{different}
\taufast \, \approx \, {\Nbarinf  \over  \vminus} {f \over (1-f)}  \comma
\gap
\tauslow \approx {\Nbarinf^2 \over 4 \vminus} \comma
                                                                \end{eq}
%-----------------------------------------------------------------------
where $1/\vminus$ is the natural time scale and $f \approx
\minf/\mtot$ is approximately the fraction of non-polymerized
monomer. Our only assumption has been that $f$ is not extremely close
to unity, \ie the system is not very close to the polymerization
temperature.  

In addition to the MWD, the other important observable is the free
monomer concentration, $\mt$.  We emphasize that since the number of
chains is fixed by the number of initiators, thus $\mt$ is linearly
coupled to $\Nbart$ by mass conservation for all times.  Hence the
relaxation of these two quantities is identical and $\mt$ relaxes
after $\taufast$.  Note that for certain other living polymers such as
end-polymerizing wormlike surfactant micelles, this is not true
\cite{marques:end-evaporation,%
milchev:monomer_mediated_relaxation,wattisking:beckerdoring}.  In
these systems there is no separate initiator species and a pair of
monomers may spontaneously unite to form a living chain. The dynamics
are thus very different and $\mt$ and $\Nbart$ may relax on different
timescales.  Marques et al. \cite{marques:end-evaporation} studied
this class of living polymer; after linearizing dynamical equations
they found that after a small $T$-jump $\mt$ relaxes much more rapidly
than $\Nbart$ which subsequently relaxes in time $\twid \Nbarinf^2$.
In numerical simulations of this same class of system, Milchev et al.
\cite{milchev:monomer_mediated_relaxation} report a $1/t$ decay of
mean chain length after a time $\twid \Nbarinf^{5}$.  In the related
system of spherical micelle aggregation
\cite{anianssonwall:living_micelle_kinetics,%
anianssonwall:living_micelle_kinetics_correction,
aniansson:micelle_kinetics_theory_expt_etal,wennerstrom:micelle_review},
the number of micelles can change and again the kinetics are very
different \cite{ben:living_ionic_letter} to the present case.

\begin{figure}[tb]             
\centering
\includegraphics[width=9cm]{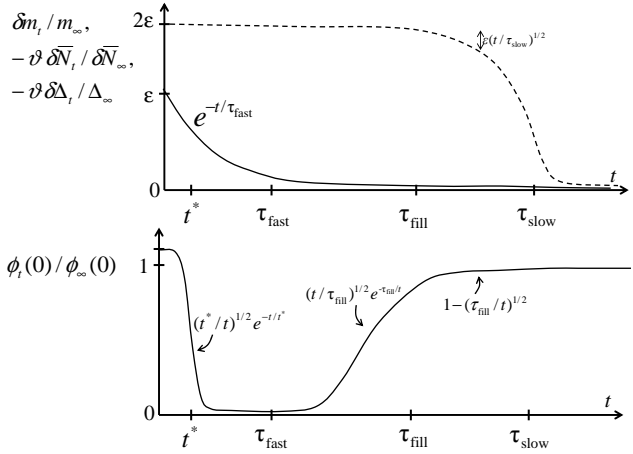}
\caption[figure]
{Time dependent recovery following a small perturbation, positive
initial boost, $\epsilon > \epsilonc$.  Top: $\delta \Nbart$, $\delta
\mt$ (solid line) become very small after $\taufast$ while the MWD dispersion
$\delta \Deltat$ (dashed line) relaxes after $\tauslow$.  Bottom: In
the process of equilibration free initiator concentration undergoes
large changes even though its initial and equilibrium values are very
close to one another.
}
\label{summary}
\end{figure}

A summary of our findings is shown in fig. \ref{summary}, for the case
emphasized here where a small $T$-jump induces a positive velocity
boost to chain growth. The figure shows the very different
relaxation times of first and second MWD moments. It also shows the
interesting behavior of the number of free initiators, $\phit(0)$,
which suffers enormous depletion at intermediate times despite
ultimately recovering to a level very close to its initial value.  We
showed that whilst virtually all of the relaxation of the fast
variables $\Nbart, \mt$ occurs on a timescale $\taufast$, they are
actually not completely relaxed by this time.  The final very late
stages of their relaxation during which their values are fine-tuned
(to relative order $1/\Nbarinf$) to the final equilibrium values,
occurs on a timescale $\tauslow$.  During this episode their values
evolve quasi-statically, enslaved to $\phit(0)$ according to $\delta
\Nbart\theta /\Nbarinf \approx -\delta \mt/\minf \approx \delta
\phit(0)$.  The novel behavior of $\phit(0)$, as well as its central
role in late relaxation of the fast variables suggests this as an
interesting quantity to measure experimentally.  This might be
achieved spectroscopically and would have the advantage of being a
relatively uninvasive probe.  We remark that for bifunctional
initiators, e.g. those used in studies by the Greer group, the meaning
of $\phit(0)$ is the concentration of ``half-chains,'' \ie those
having at least one chain end which is a bare initiator molecule.

We have demonstrated that the type of dynamical response depends on
the magnitude and sign of the perturbation. For example, if the
temperature change is reversed in sign, then the velocity boost also
changes sign.  Thus for $\alpha$-methylstyrene a negative $T$-jump
produces the positive velocity boost phenomenology of
fig. \ref{summary} whereas a small temperature increase causes a
negative boost. In this case, instead of short chains being
annihilated, their number increases dramatically during the relaxation
process.  As far as perturbation magnitude is concerned, if the
relative value $\epsilon$ is less than a critical value $\twid
1/\Nbarinf^{1/2}$ then the response is milder but remains
non-linear. For a positive boost, the hole in the MWD is no longer
complete, but instead has a relative depth less than unity but still
much bigger than $\epsilon$.

%%%%%%%%%%%%%%%%%%%%%%%%%%%%%%%%%%%%%%%%%%%%%%%%%%%%%%%%%%%%%%%%
%%%%%%%%%%%%%%%%%%%%%%%%%%%%%%%%%%%%%%%%%%%%%%%%%%%%%%%%%%%%%%%%

How do our predictions compare with experiment?  In ref.
\cite{garcia:greer:living_nonumber} small $T$-jumps were imposed on
$\alpha$-methylstyrene systems in the semi-dilute regime at
concentrations high enough to be nearly entangled.  Relaxation was
monitored by measuring viscosity $\eta$ as a function of time. Now
generally we expect viscosity to scale as a
polymer-concentration-dependent power of mean chain length, $\eta
\twid c_\phi \Nbar^{\gamma}$, where the value of $\gamma$ is predicted
by standard theories of polymer physics
\cite{doiedwards:book,gennes:book}.  For a general MWD, note the
coefficient $c_\phi$ which depends on the shape of the MWD, \ie it
depends on the full function $[\phi(N)]$. (This is because $\eta$
depends not only on the first moment, but in general has a complex
dependence on chain length distribution).  Thus we predict an initial
fast relaxation of $\eta$ lasting time $\taufast=
\Nbar/(\theta\vminus)$ (corresponding to the relaxation of $\Nbar$)
followed by a much slower relaxation in $\tauslow$ (this is the
relaxation of the prefactor $c_\phi$).  The fast relaxation time is
independent of $T$-jump magnitude (\ie independent of $\epsilon$).
Now from eqs. \eqref{v} and \eqref{rf} we can rewrite $\taufast=
1/(r\kplus\mtot)$.  Using \cite{garcia:greer:living_nonumber} $r=4.7
\times 10^{-4}$, $\mtot = 0.29 gm/cm^3$ and a measured value
\cite{zhuang:greer_kinetics} $\kplus \approx 0.2 M^{-1} sec^{-1}$ one
estimates $\taufast \approx 1000$ secs.  (Note that this work involved
bifunctional initiators.  In ref.  \cite{ben:living_ionic_letter} we
incorrectly estimated $\taufast$ to be twice the value calculated
here based on a wrong estimate of the number of living
chains which was taken to be equal to the number of bifunctional
initiators instead of twice this value.)  Similarly, with
\cite{greer:review} $f\approx 0.5$, $\vminus=0.1 sec^{-1}$ one finds
$\tauslow \approx $ 1 month.  Hence for these experiments (timescales
much less than months) $\eta$ should relax in approximately $1000$
secs.  In the experiments, jumps $\delta T\approx 1$ \degreesc were
imposed at various temperatures in the range $283^o$K $\lsim T \lsim
290^o$K.  Now the temperature dependence of $\taufast\twid 1/\kplus$
is determined by $\kplus$ which is reported
\cite{zhuang:greer_kinetics} to vary by $\approx 10\%$ over this
temperature range.  Thus we expect only a slight variation in the
viscosity relaxation time at the different temperatures studied
(despite the fact that $\Nbarinf$ changes significantly).  These
predictions are very close to the experimental findings, where
relaxation times at all temperatures were found to be about 2000
seconds, with very little variation from one temperature to another.
This is in sharp contrast to the relative changes in $\eta$ itself,
which varied by almost an order of magnitude.

We have emphasized small perturbations in this work.  In fact, the
response to a large perturbation is obtained, qualitatively speaking,
by setting $\epsilon=1$ in our results.  This case was addressed by
Miyake and Stockmayer \cite{miyakestockmayer:living} who considered
an initial condition in which all initiators are free (\ie
$\Nbar_0=0$).  They were the first to identify the two timescales of
the system, $\taufast$ and $\tauslow$, in the limit of very small
depolymerization rates ($f \ll 1$).  Their analytical results applied
to the $t < \taufast$ stage where they found an initially uniformly
translating and broadening MWD peak, reaching $\Nbarinf$ after
$\taufast$ as shown in fig. \ref{miyake_stages}.  Seen from the
viewpoint of a general perturbation of magnitude $\epsilon$, in this
case $\taufill$ and $\tauslow$ coincide because $\epsilon=1$.  That
is, the ``hole'' (in this case the entire region between $0$ and
$\Nbarinf$) fills up on the same timescale in which the whole MWD
relaxes.  The linearized dynamics, which for a small perturbation
applied for $t > \taufill$ due to $\delta \phit(0)/\phiinf(0)$
becoming small (see section 3.3) cannot be employed.  Describing
dynamics at times of order $\tauslow$ in this case is thus a difficult
nonlinear problem
\cite{nandajain:living_broadening,auluck:living_mwd_intermediate,%
taganov:living_mwd}.

In recent experiments, Greer's group
\cite{zhuang:greer_kinetics,das:greer:living_7_mwd} studied such
large perturbations using $\alpha$-methylstyrene.  After a rapid
quench below the polymerization temperature, mono\-mer concentration and
MWD were analyzed after a time delay.  The time dependent relaxation
was probed by repeating the procedure for different time delays and
samples. They found different relaxation times for the mono\-mer
concentration and the MWD width, as predicted by theory
\cite{miyakestockmayer:living}. However, $\Nbart$ and $\mt$ did not
have the same relaxation, as mass conservation would seem to dictate.
This may be due either to side-reactions during polymerization or
living chain ionic association effects
\cite{fetters:living_aggregation,stellbrink:living_aggregation,balsarafetters:living_aggregation_reeval}
which have been studied theoretically \cite{frischknechtmilner:living_aggregation}.
Dynamic light scattering measurements
\cite{ruizgarciacastillo:living_aggregation_greer} suggest that
prior to polymerization the initiators self-assemble into long
polymeric structures and it has been suggested
\cite{das:greer:living_7_mwd} that initiators may not all be
equally available for polymerization following the $T$-quench.
However, lifetimes of aggregate structures would need to be extremely
long for these effects to interfere with polymerization dynamics since
as we have seen the MWD relaxation times are very large.

In conclusion, our hope is that this work will motivate further
experimental study of the dynamical sensitivity of living polymers to
small perturbations. The drastic effect on the MWD in the small chain
region, in particular the depletion or excess of free initiators, is a
natural focus for experimental measurement.  One can think of other
{\em time-dependent} perturbations such as small amplitude thermal
cycling which would probe interesting aspects of their ultrasensitive
dynamics.  Finally, the systems we have analyzed here are model
starting points for the considerably more complex biological living
polymers, actin filaments and microtubules, which are intimately
involved in the locomotion and structural integrity of living cells
\cite{korn:actin_review_science,pollardcooper:actin_review,desaimitchison:mtub_review}.

%************************************************************************************
%************************************************************************************

\begin{acknowledgement}

This work was supported by the Petroleum Research Fund, grant no.
PRF-33944-AC7.

\end{acknowledgement}
%************************************************************************************
%************************************************************************************
%**************************** START APPENDICES **************************************
%************************************************************************************

{\appendix

\section{Derivation of Equilibrium Values $\minf$, $\Nbarinf$,
  $\vinf$, $\Dinf$.}

The equilibrium MWD can be calculated by setting the time derivative
in eq. \eqref{phi} to zero.  Its solution is the Flory distribution,
eq. \eqref{flory}, where
%_______________________________________________________________________
                                                \begin{eq}{bottle}
\Nbarinf = -\Dinf/\vinf \period
                                                                \end{eq}
%-----------------------------------------------------------------------
Solving the system of eqs. \eqref{bottle} and \eqref{mass} for
$\minf,\Nbarinf$ one has
%_______________________________________________________________________
                                                \begin{eqarray}{dump}
&& {\minf \over \mtot } = 
\inverse{2} 
\curly{1+f+{r \over 2}-\square{(1-f)^2 + r (1+3f) + {r^2 \over 4}}^{1/2}}
\drop
&& \Nbarinf = \inverse{r} \paren{1- {\minf \over \mtot}}
\period
                                                                \end{eqarray}
%-----------------------------------------------------------------------
For $r \ll 1$ and assuming $(1-f)/r^{1/2} \gg 1$, by expanding
eq. \eqref{dump} in powers of $r$ and keeping only the first term of
the expansion one recovers eq. \eqref{eqm}.  The values of
$\vinf,\Dinf$ in eq. \eqref{eqm} are obtained by substitution of
$\minf,\Nbarinf$ in eq. \eqref{vd}.

%************************************************************************************
%************************************************************************************

\section{Self-Consistency of
Results of Subsections 3.1 and 3.2}

\subsection{Coherent Chain Growth}

The solution for the $\vt$ dynamics, eq. \eqref{mi}, follows by
solving eq. \eqref{v} after neglecting the $\phit(0)$ term on the rhs.
For $t\ll\tauqs$, this is a self-consistent solution since then
the $\phit(0)$ term is negligible with respect to the remaining terms as
can be seen using the solutions for $\vt,\phit(0)$ of eqs. \eqref{mi},
\eqref{approximate}, and eq. \eqref{longphi}.

Eq. \eqref{eu} is derived by substituting $\vt$ from eq. \eqref{mi} in
eq. \eqref{phi}, replacing $\Dt = \Dinf + \delta \Dt$, and dropping
the $\delta \Dt$ term.  The fact that this term can be neglected may
be seen as follows.  Even if $\delta \Dt$ did not decrease but
remained of order $\delta D_0$ throughout this regime this would lead
to replacing $D_\infty$ by $D_\infty (1+\epsilon)$ in all results
obtained in section 3 (see eq. \eqref{beware}).  This upper bound on
the effect of $\delta \Dt$ can easily be seen to lead to higher order
in $\epsilon$ terms (see eg. eq. \eqref{approximate}) and the $\delta
\Dt$ term can therefore be neglected.

Now let us consider the propagator of the dynamics of eq. \eqref{eu},
which appears in eq. \eqref{evolve}.  For short times, $t \ll
\taufast$, one may replace $\vt$ in eq. \eqref{eu} by $v_0$.
Eq. \eqref{eu} then has constant coefficients and its propagator
including the boundary condition may be calculated exactly
\cite{hill:aggr_bio_book}:
%_______________________________________________________________________
                                                \begin{eqarray}{globe}
&& \Gtran_t (N,N') = 
\inverse{(4 \pi \Dinf t)^{1/2}} \times \drop
&&               \curly{ e^{-(N-N'- v_0 t)^2/(4 \Dinf t)}
                      + 
                        e^{-N' v_0/\Dinf}
                        e^{-(N+N'- v_0 t)^2/(4 \Dinf t)}
                      } 
               \drop
&&                 - {v_0 \over 2 \Dinf} e^{N v_0 /\Dinf}
                  \erfc \paren{N + N' +  v_0 t \over (4 \Dinf
t)^{1/2}}
\ \ \ \ \  (t \ll \taufast)
                                                                \end{eqarray}
%-----------------------------------------------------------------------
Here $\erfc(x) \equiv 1 - \erf(x)$, where $\erf$ is the error
function.  

The solution for $\phi_t(0)$ is derived by substituting
eq. \eqref{globe} in eq. \eqref{evolve} and setting $N=0$.  One has
%_______________________________________________________________________
                                                \begin{eqarray}{propro}
{\phit (0) \over \phiinf(0)} \approx 
\erfc(x) + 2 x^2 \erfc(x) - 2 \pi^{-1/2} x e^{-x^2} \comma \drop
\gap x \equiv (t/\tstar)^{1/2} 
\gap (t \ll \taufast)
                                                                \end{eqarray}
%-----------------------------------------------------------------------

Here we used the fact that since the values of $N'$ contributing to
the integration are much smaller than $\Nbarinf$, one may approximate
$\phi_0(N')$ by $\phiinf(0)$ in the integrand of eq. \eqref{evolve}
(with relative errors of order $1/\Nbarinf$).  
Eq. \eqref{approximate}
of the main text follows from eq. \eqref{propro} by taking the two
corresponding limits.  Note that eq. \eqref{propro} approximates
$\phi_t(0)$ for all $t \ll \taufast$, including times of order
$\tstar$.

\subsection{Hole Filling}

Eq. \eqref{all} is a self-consistent solution of eq. \eqref{v} since
for $t \gg \tauqs$, assuming the validity of eqs. \eqref{all} and
\eqref{approximate} one may verify that eq. \eqref{v} is satisfied
with the term on the lhs being much smaller in magnitude than both
terms on the rhs which balance one another.

Consider now the validity of eq. \eqref{simple} for $\taufast \ll t \ll
\taufill$.  Substituting eq. \eqref{longphi} into
eq. \eqref{all} one has explicit solutions for $\mt,\Nbart,\vt,\Dt$.
One thus sees that in the time regime of consideration the velocity
term in eq. \eqref{phi} is exponentially small and therefore can be
deleted in eq. \eqref{simple} with very small error.

The MWD dynamics are therefore described by eq. \eqref{simple} whose
propagator satisfying reflecting boundary conditions is
%_______________________________________________________________________
                                                \begin{eqarray}{blob}
&& \Gdiff_t (N,N') = \inverse{(4 \pi \Dinf t)^{1/2}} \times \drop
   &&            \curly{e^{-(N-N')^2/(4 \Dinf t)}
                      + 
                        e^{-(N+N')^2/(4 \Dinf t)}
                      } 
\period
                                                                \end{eqarray}
%-----------------------------------------------------------------------
The solution of eq. \eqref{simple} is thus
%_______________________________________________________________________
                                        \begin{eqarray}{bake}
\phi_t(N) & \approx & 
\int_0^\infty dN'\, \phi_{\tcross}(N') \Gdiff_{t - \tcross}(N,N')
\drop  
& \approx &
\int_{\epsilon \Nbarinf/\theta}^{\infty} dN' \,
\phi_0(N' - \epsilon \Nbarinf/\theta) \Gdiff_{t} (N,N')
\drop
&& (\taufast \ll t \ll \taufill) \period
                                                \end{eqarray}
%-----------------------------------------------------------------------
where $\tcross$ satisfies $\taufast \ll \tcross \ll \tauqs$, \ie it
belongs to the time regime described simultaneously by eqs. \eqref{eu}
and \eqref{simple}.  In going from the first integral in
eq. \eqref{bake} to the second we used the fact that at $\tcross$ the
MWD has approximately the shape of the initial MWD, boosted in the
positive direction by $\epsilon \Nbarinf/\theta
$.  The effect of diffusion
during the initial boost of the MWD may be shown to be small.  Notice
also that evolution in the last expression in eq. \eqref{bake} starts
at $t=0$; for $t \gg \tcross$ we may to leading order replace
$t-\tcross$ in $\Gdiff$ in eq. \eqref{bake} by $t$.

Substituting eq. \eqref{blob} in eq. \eqref{bake} and setting $N=0$
one obtains
%_______________________________________________________________________
                                                \begin{eq}{longfish}
{\phit (0) \over \phiinf(0)} \approx \erfc \square{\paren{\taufill \over t}^{1/2}}
\comma \ \ \ 
(\taufast \ll t \ll \taufill)
                                                                \end{eq}
%-----------------------------------------------------------------------
Considering times much shorter than the crossover time $\taufill$,
eq. \eqref{longphi} is recovered.

%************************************************************************************
%************************************************************************************

\section{Linearized Dynamics, $t \gg \taufill$}

Ultrasensitivity is such that the dynamics cannot be linearized until
the later stages, $t \gg \taufill$.  For these times we may linearize
eq. \eqref{phi} by dropping terms proportional to products of $\delta
\phit, \delta \vt, \delta \Dt$.  One has
%_______________________________________________________________________
                                                \begin{eqarray}{linearphi}
&& {\partial \delta \phit \over \partial t}
\approx
- \vinf {\partial \delta \phit \over \partial N}
+
\Dinf  {\partial^2 \delta \phit \over \partial N^2}
+
\mu_t
\comma \gap \drop
&& \mu_t \equiv {\delta \vt \over \Nbarinf} \square{\phiinf(N) - \delta(N)}
\comma
                                                                \end{eqarray}
%-----------------------------------------------------------------------
with reflecting boundary conditions at the origin.  In
eq. \eqref{linearphi} the $\delta \Dt$ term was neglected since its
magnitude is smaller by $1/\Nbarinf$ than the corresponding $\delta
\vt$ term.  The $\delta$-function in the source term in
eq. \eqref{linearphi} arises from linearizing the boundary condition
in eq. \eqref{phi}.  Notice that by definition, $\delta \phit(0)$
is normalized to zero and that the source term in
eq. \eqref{linearphi} preserves this normalization since its integral
over all $N$ is zero.

Performing the corresponding linearization in eq. \eqref{v} one has
%_______________________________________________________________________
                                                \begin{eq}{linearv}
\ddt \delta \vt
\approx
 - {1 \over \taufast} \delta \vt - {1 \over
\taufast} \Dinf \delta \phit(0)
\period
                                                                \end{eq}
%-----------------------------------------------------------------------

Now since eq. \eqref{linearphi} is of the same form as eq. \eqref{eu},
its propagator, $\Glinear_t(N,N')$, is given by eq. \eqref{globe} with
$v_0$ replaced by $\vinf$.  Thus if $\delta \phi_{t_L}(N)$ is known at time
$t_L$, then the solution of eq. \eqref{linearphi} for subsequent times
is
%_______________________________________________________________________
                                                \begin{eq}{frog}
\delta \phit(N) 
\approx 
\int_0^{\infty} dN'\, \Glinear_{t-t_L} (N,N') \delta \phi_{t_L}(N')
+
R_t(N) \comma
                                                                \end{eq}
%-----------------------------------------------------------------------
where
%_______________________________________________________________________
                                                \begin{eqarray}{m}
&& R_t(N)  
\equiv 
\int_0^{\infty} dN' \int_{t_L}^t dt' \, \Glinear_{t-t'} (N,N')
                                                \mu_{t'} (N')
\ \approx \ \drop
&& - \int_{t_L}^t dt' \, {\Dinf \delta \phit(0) \over \Nbarinf} 
                 \square{\phiinf(N) - \Glinear_{t-t'} (N,0)} \period    
                                                                \end{eqarray}
%-----------------------------------------------------------------------
Here the last expression for $R_t(N)$ is obtained by performing the
$N'$ integration using the expression for $\mu_t$ from
eq. \eqref{linearphi} and the fact that $\phiinf(N)$ remains unchanged
when evolved with $\Glinear$.  In eq. \eqref{m} we also used
%______________________________________________________________________
                                                \begin{eq}{connect}
\delta \vt \approx - \Dinf \delta \phit(0) 
                                                                \end{eq}
%-----------------------------------------------------------------------
which is valid throughout the linearized time regime as one may derive
using eq. \eqref{linearv}.  The solution of eqs. \eqref{linearphi} and
\eqref{linearv} is thus found by setting $N=0$ in eq. \eqref{frog},
solving for $\delta \phit(0)$, and then using the calculated
expression in eqs. \eqref{frog} and \eqref{connect} to find $\delta
\phit(N)$ and $\delta \vt$.  Let us now perform this analysis for two
time regimes.

(i) $\taufill \ll t \ll \tauslow$.  In this regime, we will see the
source term can in effect be ignored. The top expression on the rhs of
eq. \eqref{expo} is a self-consistent solution of eq. \eqref{frog}
which is proved as follows.  As one may see from eqs. \eqref{longphi}
and \eqref{expo} there exists a constant $\beta$ of order unity such
that for $t > \beta \taufill$ the relative perturbation in $\phit(N)$
is much smaller than unity and eq. \eqref{linearphi} applies.  (More
precisely, for any desired relative smallness there exists a different
constant $\beta$.) Thus setting $t_L = \beta \taufill$ and $N=0$ in
eq. \eqref{frog} and using the expression for $\delta \phit(0)$ from
eq. \eqref{expo} one finds that with appropriate choice of $J$,
eq. \eqref{frog} is satisfied with $R_t(0)$ being much smaller than
the remaining two terms in eq. \eqref{frog}.  The integral term in
eq. \eqref{frog} is calculated using the fact that as seen from
eqs. \eqref{longphi}, \eqref{expo}, and fig. \ref{hole_fillup},
$\delta \phi_{t_L}(N)$ is of order $\delta \phi_{t_L}(0) \gg \epsilon
\phiinf(N)$ for $N \ll \epsilon \Nbarinf/\theta$ and of order $\epsilon
\phiinf(N)$ for larger $N$. We also used
%_______________________________________________________________________
                                                \begin{eqarray}{stupid}
\Glinear_{t-t'} (0,N') 
\gt &&
\casesbracketsshortii
{1/\square{\Dinf (t - t')}^{1/2}}
                                {N' \ll \Dinf(t - t')}
{0}     
                                {N' \gg \Dinf(t - t')}
\drop && (t, t' \ll \tauslow)
                                                                \end{eqarray}
%-----------------------------------------------------------------------
which may be derived from eq. \eqref{globe}.

(ii) $t \gg \tauslow$.  Using eq. \eqref{globe} one finds
%_______________________________________________________________________
                                                \begin{eq}{red}
\Glinear_t(0,N') 
\approx
\Sinf + (\St - \Sinf) \, \lambda(N')
\ \ \  (t \gg \tauslow) \comma
                                                                \end{eq}
%-----------------------------------------------------------------------
where 
%_______________________________________________________________________
                                                \begin{eqarray}{slop}
&& \St \equiv \Glinear_t(0,0) 
\comma \
\Sinf = \phiinf(0) = 1/\Nbarinf 
\comma \drop
&& \lambda(N') \equiv 
\paren{1- {N' \over 2 \Nbarinf}} e^{N'/(2\Nbarinf)} 
\period
                                                                \end{eqarray}
%-----------------------------------------------------------------------

Now setting $t_L = \tauslow$ in eq. \eqref{frog} and using
eqs. \eqref{m} and \eqref{red} one has
%_______________________________________________________________________
                                                \begin{eqarray}{curtain}
 \delta \phit (0) \,
&& \approx 
(\St - \Sinf) \, \int_0^\infty dN' \, \lambda(N') \delta \phi_{\tauslow} (N')
+ \drop
&& \inverse{4 \Sinf \tauslow}
\int_{\tauslow}^t dt' \, \paren{S_{t-t'} - \Sinf} \delta \phi_{t'}(0)
                                                                \end{eqarray}
%-----------------------------------------------------------------------
where we used the fact that $\delta \phi_{\tauslow}(N')$ is normalized to zero. 

Now the self-consistency of the $t>\tauslow$ expression on the rhs
of eq. \eqref{expo} is proved by substituting eq. \eqref{expo} in
eq. \eqref{curtain}.  Since $\delta \phi_{\tauslow}(N')$ is of order
$\epsilon/\Nbarinf$, the magnitude of the integral of $\delta
\phi_{\tauslow}(N')$ in eq. \eqref{curtain} is of order $\epsilon$.
The last integral term in eq. \eqref{curtain} is evaluated using
eq. \eqref{expo} and
%_______________________________________________________________________
                                                \begin{eqarray}{express}
&& {\St - \Sinf} 
\, \approx \drop
&& \casesbracketsshortii
{1/(\Dinf t)^{1/2}}
                                        {t \ll \tauslow}
{{16  \Sinf \pi^{-1/2}} \paren{\tauslow/t}^{3/2} e^{-t/(16 \tauslow)} }
                                        {t \gg \tauslow} 
\drop
                                                                \end{eqarray}
%-----------------------------------------------------------------------
which may be proved using eqs. \eqref{slop} and \eqref{globe}.  One
finds that for $t \gg \tauslow$ all terms in eq. \eqref{curtain} have
the same time dependence and are of the same order of magnitude.  Thus
eq. \eqref{curtain} may be satisfied by appropriate choice of the
numerical coefficient $M$ in eq. \eqref{expo}.

%************************************************************************************
%************************************************************************************

\section{Self-Consistency of the Results of Section 5}

\subsection{Coherent Chain Shrinking}

The validity of eq. \eqref{mi} is verified using the same arguments as
in the first paragraph of Appendix B but now using the expression for
$\phi_t(0)$ from eq. \eqref{peak}.

Setting $N=0$ in eq. \eqref{bake} one obtains eq. \eqref{propro} with
$x$ now replaced by $-x$.  Considering times much greater and much
less than $\tstar$ leads to eq. \eqref{peak}.

\subsection{Peak Decay}

The validity of eq. \eqref{all} is proved  using exactly the same
arguments as the ones in the first paragraph of the corresponding
section of Appendix B, but now using the expression for $\phi_t(0)$
from eq. \eqref{phispread}.  One finds that the crossover time $\tauqs
\approx \taufast \ln (\Nbarinf)$ is different from the respective
crossover time found in Appendix B.

The solution of eq. \eqref{simple} is now
%_______________________________________________________________________
                                                 \begin{eqarray}{cake}
 \phit(N) \approx \int_0^{\infty} dN' \,
 \phi_0(N' - \epsilon \Nbarinf/\theta) \Gdiff_t(N,N')
 \ + \drop
 \Gdiff_t (N,0)  \int_0^{\epsilon \Nbarinf/\theta} dN' \, \phi_0(N') 
 \ \ \                                  
 (\taufast \ll t \ll \tauslow) \period
                                                                 \end{eqarray}
%-----------------------------------------------------------------------
Here the evolved MWD consists of two parts, corresponding to the two
terms on the rhs (see fig. \ref{trans_strong_peak}(c)).  The first is
the initial MWD, $\phi_0$, shifted by $\epsilon \Nbarinf/\theta$
towards smaller $N$.  The second part is the excess peak accumulated
by the negative boost at the origin and whose amount was the polymer
initially lying between zero and $\epsilon \Nbarinf/\theta$.  Here, as
in section 3, the effects of diffusive broadening during $t <
\taufast$ have been neglected, and $\taufast$ in eq. \eqref{cake} has
been replaced by zero.

Now substituting eq. \eqref{blob} in eq. \eqref{cake} and setting
$N=0$ one obtains eq. \eqref{phispread} similarly to the derivation of
eq. \eqref{longphi} in Appendix B.

%************************************************************************************
%************************************************************************************

\section{Self-Consistency of the Results of Section 6}

For times $t \ll \taufast$, all results and analysis are identical to
those for the strong perturbation ($\epsilon > \epsilonc$) case in the
regime $t \ll \tstar$.  Eq. \eqref{weak-all} can be shown to be a
self-consistent solution using the same arguments as those of
Appendices B and D to prove eqs. \eqref{mi} and \eqref{all},
respectively.  For times longer than $\taufast$ linearization of the
dynamics can be performed and the results of Appendix C directly
apply.

%*************************************************************************************
%*************************************************************************************
%******************************* END APPENDICES **************************************
%*************************************************************************************

}

%****************************** BIBLIOGRAPHY *****************************************

%\bibliography{polreaction,allreaction,polymerization,polgeneral,ben_dimitris,ben,rgcritical,radical,living,living_dimitris,actin_tubulin,bio_general,polgeneral_dimitris,living_note}

\end{document}